\documentclass[12pt]{article}
\usepackage{epsf}
\usepackage{amsmath}
\usepackage{amsfonts}
\usepackage{amssymb}
\usepackage{graphicx}
\usepackage{color}
\usepackage{psfrag}
\usepackage{cite}
\usepackage{hyperref}
\usepackage{tikz}
\usepackage{ifpdf}

\newcommand{\bmat}{\left(\begin{array}}
\newcommand{\emat}{\end{array}\right)}

\def\yzero{\smash{\hbox{$y\kern-4pt\raise1pt\hbox{${}^\circ$}$}}}

\def\beq{\begin{equation}}
\def\eeq{\end{equation}}
\def\beqa{\begin{eqnarray}}
\def\eeqa{\end{eqnarray}}

\def\-{\hphantom{-}}

\def\s2{\frac{1}{\sqrt2}}

\def\beq{\begin{equation}}
\def\eeq{\end{equation}}
\def\beqa{\begin{eqnarray}}
\def\eeqa{\end{eqnarray}}

\def\IF{\relax{\rm I\kern-.18em F}}
\def\II{\relax{\rm I\kern-.18em I}}

\def\Dsl{\,\raise.15ex\hbox{/}\mkern-13.5mu D} %this one can be subscripted

\def\IZ{{\bf Z}}
\def\IX{{\bf X}}

\newcommand{\eq}[1]{(\ref{#1})}
\newcommand{\ket}[1]{\vert #1 \rangle}

\catcode`\@=11   
\newdimen\@rotdimen
\newbox\@rotbox  

\def\@vspec#1{\special{ps:#1}}%  passes #1 verbatim to the output
\def\@rotstart#1{\@vspec{gsave currentpoint currentpoint translate
   #1 neg exch neg exch translate}}% #1 can be any origin-fixing transformation
\def\@rotfinish{\@vspec{currentpoint grestore moveto}}% gets back in synch 
\def\@rotr#1{\@rotdimen=\ht#1\advance\@rotdimen by\dp#1%
   \hbox to\@rotdimen{\hskip\ht#1\vbox to\wd#1{\@rotstart{90 rotate}%
   \box#1\vss}\hss}\@rotfinish}
\def\@rotl#1{\@rotdimen=\ht#1\advance\@rotdimen by\dp#1%
   \hbox to\@rotdimen{\vbox to\wd#1{\vskip\wd#1\@rotstart{270 rotate}%
   \box#1\vss}\hss}\@rotfinish}%
\def\@rotu#1{\@rotdimen=\ht#1\advance\@rotdimen by\dp#1%
   \hbox to\wd#1{\hskip\wd#1\vbox to\@rotdimen{\vskip\@rotdimen
   \@rotstart{-1 dup scale}\box#1\vss}\hss}\@rotfinish}%
\def\@rotf#1{\hbox to\wd#1{\hskip\wd#1\@rotstart{-1 1 scale}%
   \box#1\hss}\@rotfinish}%
\def\rotate{\@ifnextchar[{\@rotate}{\@rotate[l]}}
\def\@rotate[#1]#2{\setbox\@rotbox=\hbox{#2}\@nameuse{@rot#1}\@rotbox}

\catcode`\@=12
\topmargin
-1.5cm
\textwidth
15.5cm
\textheight
23.5cm
\oddsidemargin
0.7cm
\evensidemargin
1.2cm

\begin{document}

%----------------------------------------------------------------------%
%  numbering equations with section number
%----------------------------------------------------------------------%
\makeatletter
\@addtoreset{equation}{section}
\makeatother
\renewcommand{\theequation}{\thesection.\arabic{equation}}
%----------------------------------------------------------------------%
%  title page
%----------------------------------------------------------------------%
\pagestyle{empty}
%\vspace*{1.0in}
\rightline{ IFT-UAM/CSIC-15-028}
 \rightline{ FTUAM-15-8}
\vspace{1.2cm}
\begin{center}
\LARGE{\bf Transplanckian axions !? \\[12mm]}
\large{Miguel Montero$^{1,2}$,  Angel M. Uranga$^2$, Irene Valenzuela$^{1,2}$\\[4mm]}
\footnotesize{${}^{1}$ Departamento de F\'{\i}sica Te\'orica, Facultad de Ciencias\\[-0.3em] 
Universidad Aut\'onoma de Madrid, 28049 Madrid, Spain\\
${}^2$ Instituto de F\'{\i}sica Te\'orica IFT-UAM/CSIC,\\[-0.3em] 
%C/ Nicol\'as Cabrera 13-15, 
Universidad Aut\'onoma de Madrid, 28049 Madrid, Spain} \\ 

\vspace*{2.5cm}

\small{\bf Abstract} \\[5mm]
\end{center}
\begin{center}
\begin{minipage}[h]{16.0cm}
We discuss quantum gravitational effects in Einstein theory coupled to periodic axion scalars to analyze the viability of several proposals to achieve superplanckian axion periods (aka decay constants) and their possible application to large field inflation models. The effects we study correspond to the nucleation of euclidean gravitational instantons charged under the axion, and our results are essentially compatible with (but independent of) the Weak Gravity Conjecture, as follows: Single axion theories with superplanckian periods contain gravitational instantons inducing sizable higher harmonics in the axion potential, which spoil superplanckian inflaton field range. A similar result holds for multi-axion models with lattice alignment (like the Kim-Nilles-Peloso model). Finally, theories with $N$ axions can still achieve a moderately superplanckian periodicity (by a $\sqrt{N}$ factor) with no higher harmonics in the axion potential. The Weak Gravity Conjecture fails to hold in this case due to the absence of some instantons, which are forbidden by a discrete $\IZ_N$ gauge symmetry. Finally we discuss the realization of these instantons as euclidean D-branes in string compactifications.
\end{minipage}
\end{center}
\newpage
%----------------------------------------------------------------------%
%  Resetting of counters
%----------------------------------------------------------------------%
\setcounter{page}{1}
\pagestyle{plain}
\renewcommand{\thefootnote}{\arabic{footnote}}
\setcounter{footnote}{0}
%----------------------------------------------------------------------%
%  Paper begins
%----------------------------------------------------------------------%

\tableofcontents

\vspace*{1cm}

\section{Introduction}

The present precision era in the measurement of CMB observables is triggering a vigorous activity in model building of early universe inflationary models (see \cite{Baumann:2014nda} for a recent string-motivated review). 
The most recent example is provided by the possibility of a sizable tensor to scalar ratio $r$, raised by the initial BICEP2 detection claim of B-mode polarization from primordial gravitational waves \cite{Ade:2014xna}, and still allowed by the joint Planck/BICEP2 upper bound $r<0.12$ \cite{Ade:2015tva}). Due to the correlation, in single field inflation models, of this ratio and the inflaton field range by the Lyth bound \cite{Lyth:1996im}, there has been an intense activity in building  inflation models in which the inflaton travels transplanckian distances in field space. 

Such  large-field inflation models are sensitive to an infinite number of corrections to the inflaton potential even if they are suppressed by the Planck mass scale, so their construction requires accounting for corrections due even to quantum gravity effects. Given the difficulties in addressing the latter, a sensible approach is to invoke symmetries protecting the model against such corrections; therefore, many large-field inflation models are based on axions, i.e. scalars $\phi$ with an approximate continuous shift symmetry, broken by non-perturbative effects $e^{i\phi/f}$ to a discrete periodicity
\beqa
\phi\sim \phi +2\pi f.
\eeqa
This idea was originally proposed at the phenomenological level in so-called natural inflation  \cite{Freese:1990rb}, assuming a cosine gauge instanton potential, and requiring axions with  transplanckian decay constant $f$ (see \cite{Croon:2014dma} for recently suggested exceptions). The `empirical' case-by-case realization that this requirement is not obviously realized in string theory compactifications \cite{Banks:2003sx} (which provide a template for a quantum gravity framework), motivated the construction of models with multiple axions and potentials from gauge instantons, in which some axion linear combination effectively hosts a transplanckian field range in its basic period, even if the original periodicities are taken sub-planckian \cite{Kim:2004rp,Dimopoulos:2005ac} (see \cite{Choi:2014rja,Higaki:2014pja,Tye:2014tja,Kappl:2014lra,Bachlechner:2014hsa,Ben-Dayan:2014zsa,Higaki:2014mwa,Bachlechner:2014gfa,Burgess:2014oma,Gao:2014uha} for recent works).

These models are formulated in purely phenomenological field theory terms, and therefore there remains the  question of whether their features survive in actual embeddings in theories including quantum gravitational corrections. In fact, there are indirect arguments (under the name of the Weak Gravity Conjecture (WGC) \cite{ArkaniHamed:2006dz}) for the existence of additional instanton effects leading to higher harmonics in the axion potential, i.e. $e^{in\phi}$, in any consistent theory of quantum gravity (see \cite{Rudelius:2014wla,Rudelius:2015xta} for similar work). Because the presence of gravity is essential in the argument, these contributions may be beyond those captured by gauge instantons in the phenomenological model. If sufficiently strong, such contributions could spoil the transplanckian field range by introducing additional maxima along the axion period. Unfortunately, the WGC arguments are insufficient to quantify the strength of these effects. 

In this paper we undertake a direct approach to this problem, by considering euclidean Einstein gravity and explicitly constructing gravitational instantons coupling to the relevant axions, and evaluating their contributions to the axion potential. The configurations are wormhole solutions, and for sufficiently large axion charge have low curvature and should provide good effective descriptions of the corresponding configuration in any UV quantum theory containing Einstein gravity. Our results can be summarized as follows:

$\bullet$ For single axion models, the charge-$n$ gravitational instanton action is suppresed by the axion period $f$ as 
\beqa
S_E\sim \frac{nM_P}{f}
\eeqa
Hence, axions with parametrically transplanckian period receive unsuppressed higher harmonic contributions to their potential from gravitational instantons. These generically prevent achieving transplanckian inflationary field ranges.

$\bullet$ For multiple axion models, effective transplanckian axion periods can in principle be achieved for certain axion linear combinations.
However for models based on lattice alignment \cite{Kim:2004rp}, parametrically large field ranges also receive unsuppressed higher harmonic contributions and the results are similar to the single axion case: gravitational instantons jeopardize inflationary transplanckian field ranges. 

$\bullet$ Finally, for theories with $N$ axions, it is still possible to achieve a  superplanckian periodicity (by an enhancement of a $\sqrt{N}$ factor) corresponding to travelling along a diagonal in the field space, with no higher harmonics in the axion potential. The absence of the missing instantons is due to a $\IZ_N$ discrete gauge symmetry, reflected in the existence of certain $\IZ_N$ strings. We show that this result can be used to generalize the Weak Gravity Conjecture by including the effects of a $\IZ_N$ discrete gauge symmetry.

\medskip

The gravitational instantons are effective descriptions of configurations in concrete models of quantum gravity like string theory. For instance, in realizations of natural or aligned inflation, similar effects can arise from euclidean D-brane instantons beyond those corresponding to gauge instantons (dubbed exotic or stringy \cite{Blumenhagen:2006xt,Ibanez:2006da,Florea:2006si}, see \cite{Blumenhagen:2009qh,Ibanez:2012zz} for a review). An important aspect, which we emphasize and has seemingly been overlooked in the literature
(see \cite{Grimm:2007hs,Ben-Dayan:2014lca,Long:2014dta,Shiu:2015uva,Kappl:2015pxa,Shiu:2015xda} for recent attempts to embed such models in string theory), is that the corresponding instantons may not correspond to BPS instantons in the vacuum. Non-BPS instantons are not often discussed in the literature, because in supersymmetric setups they do not lead to corrections to the superpotential, but rather to higher F-terms (higher derivative or multi-fermion terms) \cite{GarciaEtxebarria:2008pi}. However, in non-supersymmetric situations like inflation, their extra fermion zero modes are lifted or equivalently the external legs are saturated (by insertions of susy breaking operators), and their contributions descend to the scalar potential \cite{Uranga:2008nh}. These considerations should be implemented to analyze actual string theory realizations of natural and aligned inflation (see \cite{Kenton:2014gma} for a possible exception).

Before concluding, let us mention an alternative to natural or aligned inflation, the so-called axion monodromy inflation, which uses a single axion with subplanckian period, but with
 a multivalued potential (and monodromy arising either via brane couplings \cite{Silverstein:2008sg,McAllister:2008hb,Berg:2009tg,Palti:2014kza} or via potentials from flux backgrounds \cite{Marchesano:2014mla,Blumenhagen:2014gta,Hebecker:2014eua,McAllister:2014mpa,Garcia-Etxebarria:2014wla}, see also \cite{Ibanez:2014kia,Franco:2014hsa,Blumenhagen:2014nba,Hebecker:2014kva,Ibanez:2014swa
}), see  \cite{Kaloper:2008fb,Kaloper:2011jz,Kaloper:2014zba} for a 4d phenomenological approach. For the purposes of the present paper, the subplanckian nature of the axion period suffices to protect these models against the effects described in this paper. In any event, gravitational instantons would still give rise to periodic modulations of the potential which could lead to detectable signatures in the primordial power spectrum \cite{Flauger:2009ab}.

The recent work \cite{Rudelius:2015xta} also investigates the consequences of the Weak Gravity Conjecture for axion inflationary models in string theory along lines similar to our own. 

The paper is organized as follows. In section \ref{sec:wgc} we review the Weak Gravity Conjecture, and its application to axion scalars. In section \ref{single} we describe the quantum gravitational corrections to the scalar potential in single axion models (section \ref{sec:instanton}), and its implications for transplanckian axions (section \ref{sec:impli-single}). In section \ref{multiple} we extend the analysis to multi-axion models. We discuss different alignment mechanisms in section \ref{sec:align}, and explore the implications of gravitational corrections in section \ref{sec:impli-multi}. Section \ref{twomore} reconciles the appearance of moderately superplanckian axion periods in certain alignment models with the Weak Gravity Conjecture. The systems contain charged strings, discussed in section \ref{sec:charged-string}, related to discrete gauge symmetries which imply a generalization of the WGC, as discussed in section \ref{sec:putting}. Section \ref{sec:string} discusses some aspects of single and multiple axion models in string theory, including some general considerations (section \ref{string-gen}), the relation between D-brane instantons and gravitational instantons (section \ref{string-match}) and applications to aligned axion models (section \ref{string-align}). Finally, section \ref{conclu} contains our conclusions.

\section{Remarks on the Weak Gravity Conjecture}
\label{sec:wgc}

We will study the possibility of transplanckian field ranges in a quantum theory of gravity from a semiclassical perspective, focusing on those effects which can be safely described using the Einstein-Hilbert action for the gravitational field. Although our approach is independent of the validity of the Weak Gravity Conjecture \cite{ArkaniHamed:2006dz}, its rationale is at the heart of our approach.

The essence of the WGC can be reduced to the statement that, for any abelian $p$-form field, there must be a charged $p$-dimensional object with tension
\begin{align}T\lesssim \frac{g}{\sqrt{G_N}}.\label{wgc}\end{align}
Here, $g$ is the coupling of the $p$-form field to its sources, and it has dimensions of $[\text{mass}]^{p+1-D/2}$. The rationale for the conjecture is that one expects to be able to build black $p-1$ membranes electrically charged under the abelian $p$-form field. These black branes will generically evaporate via Hawking radiation, radiating their charge away. However, this is only possible if there is an object lighter than the black brane and with a smaller charge. The extremal black hole has tension given by the right hand side of \eq{wgc}, and hence the constraint. Otherwise we would end up with a large number of Planck-sized remnants, with the associated trouble \cite{Susskind:1995da}. There are two posssible loopholes to the above conjecture, which we now discuss in some detail.

\subsection{Dependence on the spectrum of the theory} \label{depespec}

The bound \eq{wgc} comes from the tension of an extremal black brane of Planckian size. To determine this tension is a very nontrivial question which depends on the specific theory under consideration, and it is possible to find theories in which the tension of the extremal black hole is larger than the right hand side of \eq{wgc}. For instance, black branes could be charged under $\IZ_N$ discrete gauge symmetries \cite{Banks:2010zn} which effectively multiply the coupling $g$ by $N$, so that the right WGC bound would be $N$ times \eq{wgc}. We will see an example of this in section \ref{twomore}, in the context of instantons and $0$-forms, but let us advance the main idea in the usual WGC setup of $U(1)$ gauge theory in 4d.

Consider a theory with two $U(1)$'s in 4 dimensions, with lagrangian
\begin{align}\mathcal{L}=\frac12\mathcal{G}_{ij}F^i\wedge * F^j\end{align}
The metric $\mathcal{G}$ encodes the possibility of kinetic mixing between the two $U(1)$'s. In this theory, we will be able to build Reissner-Nordstrom black holes charged under both $U(1)$'s. The extremal black hole with charges $Q_1$, $Q_2$ (which we take as integers in order to comply with Dirac quantization) has mass
\begin{align}M_{\text{extremal}}=M_P\sqrt{\mathcal{G}_{ij}^{-1}Q_iQ_j}.\label{wgb}\end{align}
The WGC would demand the existence of charged particles with a charge-to mass ratio smaller than that of the above extremal black holes, for any values of $(Q_1,Q_2)$. 

Now, consider a degenerate limit in which $\mathcal{G}$ has rank one. For instance, take
\begin{align}\mathcal{G}=\frac{1}{2g^2}\left(\begin{array}{cc}1&1\\1&1\end{array}\right).\label{kinsing}\end{align}

That is, only one particular linear combination of the two $U(1)$'s, namely
\begin{align}A_{+}=\frac12(A_1+A_2)\end{align}
is dynamical. The normalization has been chosen so that black holes have integer charges. The kinetic term is just $\frac{1}{g^2}F_+\wedge *F_+$. For all intends and purposes, it seems we have a theory with just a single $U(1)$ with coupling $g/\sqrt{2}$. The WGC would seem to give us a bound $M_{\text{extremal}}\sim M_P g/\sqrt{2}$, that is, a particle with charge 1 and lighter than this should exist. This is however not correct, as we will now explain.

 The fact that the antidiagonal gauge field $A_{-}$ is nondynamical does not mean it has no effects on the physics. On the contrary, in any theory with charged matter (which in our case would correspond to the black holes), there is a coupling of the form $A\wedge *j$ in the action. If the gauge field is nondynamical, it acts as a Lagrange multiplier enforcing the condition $j=0$. In other words, only states with no charge under $A$ are admissible.

For a black hole with charges $(Q_1,Q_2)$, the charges under the diagonal and antidiagonal combinations are $Q_\pm=(Q_1\pm Q_2)$.  Requiring that $Q_-=0$ amounts to saying that $Q_+$ is even. In other words, the effect of the antidiagonal $U(1)$ is to add an extra selection rule which removes all the oddly charged black holes from the spectrum. 

This has dramatic consequences for the bound of the WGC. There is no physical reason to impose that the black hole of charge 1 must be able to decay if said black hole is not in the spectrum to begin with. The best we can do is to repeat the argument with the black hole of charge 2, the first one known for sure to be in the theory. As a result, we get a bound twice as large, i.e. $M\lesssim 2M_{\text{extremal}}\sim M_P \sqrt{2}g$. That this is the right bound can also be seen also by looking at \eqref{wgb} with $\mathcal{G}$ given by \eqref{kinsing}. Indeed,  \eqref{wgb} is only nontrivial for black holes charged along the diagonal $U(1)$, that is, with $Q_1=Q_2$. For the smallest black hole with $Q_1=Q_2=1$, \eqref{wgb} gives again $M_P \sqrt{2}g$. Notice that the generalized Weak Gravity Conjecture of \cite{Cheung:2014vva} (see also \cite{Rudelius:2014wla,Rudelius:2015xta}) would have resulted in the more stringent bound $M\lesssim M_P g/\sqrt{2}$, which as discussed above does not hold. 

Another way to rephrase the above is that the nondynamical antidiagonal $U(1)$ implements a $\IZ_2$ discrete gauge symmetry. Since the allowed charged matter is of even charge, global gauge transformations of the form $\theta\rightarrow \theta+\pi$ constitute a symmetry of the theory, even of its nonperturbative sector.

Of course, all of the above discussion can be rephrased as a redefinition of the gauge coupling/charge normalization. One may rescale a $U(1)$ gauge field by an arbitrary factor,  $A\rightarrow \alpha A$. This changes the coupling $\alpha\rightarrow g/\alpha$ and the charge quantum (which can be an arbitrary real number) gets multiplied by $\alpha$. There is a unique choice of $\alpha$ for which the charge quantum is 1, i.e. the charge takes integer values. What we have just seen is that with coupling $g/\sqrt{2}$ the charge can take only even values, i.e. the charge quantum is 2. This is equivalent to a theory in which the charge quantum is 1 and the gauge coupling is $g\sqrt{2}$. Applying the usual WGC bound then yields the right result. Our main point is that in order to apply these bounds correctly, knowledge of the spectrum of allowed objects in the theory is essential. 

\subsection{Weak gravity conjecture and axions}
 For $p=0$, that is, an axion, the coupling $g$ is nothing but the inverse of the axion decay constant $f$, and so the above conjecture formally implies the existence of an instanton with action $S\lesssim \frac{M_P}{f}$. This is usually taken to imply \cite{ArkaniHamed:2006dz} that the axion cannot have a parametrically flat potential. However, there is a very strong difference between $p=0$ and $p>0$; whereas in the latter we do have stable black hole (in general, black brane) remnants unless an object satisfying \eq{wgc} exists, there is no similar argument for the axion, since there is no would be stable black hole solution: the electrically charged object is a gravitational instanton. When $f\gg M_P$, we have many such instantons with small action, and computing their effect will be a challenge; however, there does not seem to be any inconsistency of the kind described in \cite{ArkaniHamed:2006dz}.
 
Although this seems to suggest that the WGC for axion fields is not on such a firm ground, it does highlight the fact that in a consistent quantum theory of gravity one expects to have gravitational instantons coupled to the axion. These can be found just by solving the Euclidean equations of motion for the axion-gravity system, while specifying the right asymptotic charge. This is in accordance with the considerations in \cite{Banks:2010zn} that in a quantum theory of gravity one expects to have all possible charged objects. 

As an example, consider Maxwell-Einstein theory with no charges. One can produce, in a very strong electric field, pairs of Reissner-Nordstrom black holes\cite{PhysRevD.49.958,Hawking:1994ii,PhysRevD.49.2909}. Therefore, the fact that the theory has nonsingular configurations with nonvanishing charge, together with the quantum pair-production process, should ensure the existence of these objects in the full theory.

A similar argument works also for Euclidean theories. As discussed above, one can build a solution to the euclidean equations of motion with nontrivial axion charge. This is an instanton. The question now is wether it is possible to build a consistent theory in which these instantons are excluded of the path integral, or on the contrary their presence cannot be avoided. This seems unlikely, since in a quantum theory of gravity we are expected to add up all geometries with the right asymptotics \cite{1993AdSAC...8..163H}. Although a single instanton does not fulfill this condition, since its axionic electric charge may be measured at infinity, an instanton- anti instanton pair does. Once we allow the instanton- anti instanton amplitude in the path integral, it is difficult to exclude the single instanton as a configuration mediating a transition between two different states; one could always cut a time slice between the instanton and the anti-instanton. This constant time configuration defines a state within the theory, so it makes sense to consider finite action configurations connecting it to the vacuum (that is, a single instanton). 

Probably, the smallest such instanton will involve Planckian geometries, and so it should not be taken very seriously. However, some instantons may turn out to be within the reach of semiclassical Einstein gravity.  Their effects on axion dynamics can be safely studied even in the absence of an UV completion.

\section{Gravitational instantons for a single axion\label{single}}

In this section we start constructing gravitational instantons coupling to axions, and their effects, in theories of gravity coupled to axions. These do not include other fields like scalar partners of the axions to turn them into complex fields, fermions or other superpartners, etc. The motivation for this is twofold: First, for applications to single-field inflation, the inflationary dynamics requires the axion to be the only (non-gravitational) dynamical field; in fact, the main challenge in realizing inflation in UV complete theories, like string theory, lies in giving hierarchically large masses to all fields except for the inflaton, so that they do not interfere with inflation. Hence, gravity coupled to axions provides the appropriate setup to address questions pertaining the possible gravitational corrections to the flatness of the inflationary potential. Second, we would like to keep our framework as minimal and general as possible, to explore the interplay of gravitational instantons, axions, and the Weak Gravity Conjecture in general gravitational theories, in a model-independent way.

Throughout the paper we assume that the axion which is to act as the inflaton somehow has an adequate potential to provide successful inflation. In typical models this comes e.g. from a gauge sector coupled to the axion. We will not address in this paper how this potential is obtained: we will take it as given and then ask ourselves if gravitational effects can spoil it. 

We first study the consequences of gravitational instantons for a single axion with periodicity $\phi\sim\phi+2\pi$. The Minkowskian action for the theory is 
\begin{align}\int\left(\frac{-1}{16\pi G}R dV+ \frac{f^2}{2} d\phi\wedge *d\phi\right)+\ldots\label{action1inst}\end{align}
where the dots stand for any extra light fields. The euclidean counterpart of \eq{action1inst} is
\begin{align}\int \left(\frac{-1}{16\pi G}R dV- \frac{f^2}{2} d\phi\wedge *d\phi\right)+\ldots\label{action1instE}\end{align}
Notice that naive Wick rotation would have produced the opposite kinetic term for the axion. This extra minus sign for an axion as compared to a standard scalar field is related to it being dual to a three-form; duality involves the Hodge star operator and since $*^2=(-1)^{k(n-k)}s$ for $k$-forms in $n$-dimensional space with metric signature $s$, dualizing to a three-form and Euclidean rotation do not commute. The best semiclassical approximation is obtained by first dualizing to a three-form, then performing euclidean rotation, and then dualizing back to a scalar; as a result the extra minus sign arises. For a detailed explanation of this point, see \cite{ArkaniHamed:2007js}.

\subsection{The instanton}
\label{sec:instanton}

We are looking for a solution of \eq{action1instE} which is asymptotically flat and with an axion profile which at large distances from its core has the asymptotic form
\begin{align}\phi\sim\frac{n}{4\pi^2f^2}\frac{1}{r^2}.\end{align}
This means it has electric charge $n$, since asymptotically $*d\phi=\frac{n}{2\pi^2f^2}d\Omega_ 3$. 

With everything discussed above, we only need to find the right gravitational instanton. This will be a spherically symmetric solution with the right axionic charge. To do this we take a spherically symmetric ansatz for the metric,
\begin{align} ds^2=f(r)dr^2+r^2ds^2_{S^3},\end{align}
where we have redefined the radial coordinate so that constant-$r$ surfaces behave like proper spheres. The only dependence of the metric is through the radial factor $f(r)$. Similarly, the axion profile will be taken as a function only of $r$. We thus get a system of two coupled second-order ODE's which is easy to analyze in detail.

The stress-energy tensor of an axion field is
\begin{align}f^{-2}T_{\mu\nu}=\partial_\mu\phi\partial_\nu\phi-\frac12(\partial_\mu\phi\partial^\mu\phi)g_{\mu\nu}.\end{align}
The only nonvanishing components of these are
\begin{align}T_{rr}=\frac12 f^2(\phi')^2,\quad T_{ii}=-\frac{1}{2f}(f\phi')^2g_{ii},\ \ i=\psi,\theta,\phi.\end{align}

Einstein's equations yield 
\begin{align}\frac{3}{r^2}\left(1-f\right)&=-4\pi G(f\phi')^2,\nonumber\\-f+1-r\frac{f'}{f}&=4\pi Gr^2(f\phi')^2.\end{align}
The solution is
\begin{align}f(r)=\frac{1}{1-\frac{a}{r^4}},\quad d\phi(r)= \frac{n}{2\pi^2f^2}\sqrt{f(r)}\frac{dr}{r^3},\quad a\equiv\frac{n^2}{3\pi^3}\frac{G}{f^2}=\frac{n^2}{3\pi^3}\left(\frac{M_P}{f}\right)^2M_P^{-4}.\end{align} %\sqrt{\frac{3f(r)}{2\pi G f^2a}}
The parameter $a^{1/4}$ gives the typical size of the instanton; at this radius the metric seems to become singular. This is a coordinate singularity, as the curvature scalar 

\begin{align}R=8\pi G T=8\pi G f^2 \partial_\mu\phi\partial^\mu\phi=8\pi \frac{G}{f^2} \left(\frac{n}{4\pi^2}\right)^2\frac{1}{r^6}=\frac32 \frac{a}{r^6}.\end{align}
 is perfectly regular there. Notice that for $r=a^{1/4}$, which corresponds to the maximum curvature at the throat of the wormhole, $R=\frac32 a^{-1/2}$. Thus the single parameter controlling wether the solution can be trusted is $a$; it has to be larger than the Planck scale for the solution to be reliable within the context of effective field theory. Notice also that the wormhole radius $\sim a^{1/4}$ is of order Planck for $n=1$; we need wormholes with higher axionic charge.
 
 Since we have defined our radial coordinate in terms of the areas of 3-spheres, what is happening is that as we go to $r=a$ from infinity these areas reach a minimum, then start growing again. In fact, by making the change of coordinates $r^4=a+t^2$, the metric becomes
\begin{align}ds^2=r^2\left[\frac14dt^2+ds^2_{S_3}\right],\end{align}
which is conformally equivalent to a flat metric on $S^3\times \mathbb{R}$; in other words, our instanton is actually a wormhole. In fact, we have just rederived the Strominger-Giddings wormhole \cite{Giddings:1987cg}. As discussed there, it is up to us to choose which geometries is the wormhole connecting: Two asymptotically flat regions is clearly unphysical. We might think of an euclidean wormhole connecting two points of spacetime, with an arbitrarily long tube between them; thus, we might better think of half our solution describing the ``entrance'' of a wormhole (instanton), and a solution with negative axionic charge describing the exit (anti-instanton). 

In \cite{Giddings:1987cg}, these half-wormholes are regarded as describing the nucleation of baby universes, which then evolve on their own towards a Big Crunch.  The interpretation is very similar to a gauge instanton, which changes the asymptotic configuration of the gauge fields, in particular  the Chern-Simons number of the vaccum
\begin{align}n_{CS}=\frac13\int_{\mathbb{R}^3}\mathbf{A}\wedge\mathbf{A}\wedge\mathbf{A}.\end{align}

Since two states with different $n_{CS}$ are degenerate in energy, the true vacuum is a $\ket{\theta}$ vacuum of the form
\begin{align}\ket{\theta}=\sum_{n_{CS}} e^{in_{CS}\theta} \ket{n_{CS}}.\end{align}

The analog of the Chern-Simons number in the gravitational case would be the number of connected components of space. A half-wormhole nucleates a baby universe which undergoes a Big Crunch; after the wormhole has closed, we end up with a Universe with two connected components. The instanton thus increases this number by one. Again by analogy with the gauge case, one could think that the actual ground state is one of the so-called ``$\alpha$-states'' of Coleman \cite{PhysRevLett.52.1733,Coleman:1988cy,Giddings:1988cx}; in this case, the half-wormholes would generate a potential for the axion in exactly the same way as the instantons in gauge theory. We will discuss the effect of the instanton in the axion potential further below.

The only other relevant quantity of the solution is its action. This can be computed by noting that the equations of motion imply $R=8\pi G T$, and $T=-f^2(\partial \phi)^2$, so that\footnote{We thank \cite{Bachlechner:2015qja} for pointing out a missing factor of $1/2$ in an earlier version of this manuscript.}
\begin{align}
S_E=\frac{f^2}{2}\int dV\ \partial_\mu \phi\partial^\mu\phi=\pi^2f^2 \int_{a^{1/4}}^\infty f(r)^{-1/2}r^3 (\phi'(r))^2\ dr=\frac{\sqrt{3\pi}}{16}\frac{M_P n}{f}
\label{instac1}.
\end{align}
where in the first equality we have divided by two to take into account the fact that our instanton is only half a wormhole.

We now have to argue that this gravitational instanton can indeed generate the right potential for the axion. Let us expand the axion field in the background of an instanton of charge $n$ as $\phi\approx\langle\phi_n\rangle+\varphi$, where $\varphi$ is the fluctuation. The axion kinetic term thus yields
\begin{align}&d(\langle\phi_n\rangle+\varphi)\wedge *d(\langle\phi_n\rangle+\varphi)\nonumber\\&=d\langle\phi_n\rangle\wedge *d\langle\phi_n\rangle+d\varphi\wedge *d\varphi+d\langle\phi_n\rangle\wedge* d\varphi+d\varphi\wedge*d\langle\phi_n\rangle.\end{align}
The last two terms may be rearranged as
\begin{align}d\langle\phi_n\rangle\wedge* d\varphi+d\varphi\wedge*d\langle\phi_n\rangle&=2d\varphi\wedge*d\langle\phi_n\rangle=2d(\varphi\wedge *d\ \langle\phi_n\rangle)-2\varphi\wedge (d*d\langle\phi_n\rangle)\nonumber\\&\sim-2n\varphi\wedge (d*d\langle\phi_n\rangle) ,\end{align}
where we dropped the total derivative term. We thus get the usual axion-instanton coupling term, with the instanton supported on the form $(d*d\langle\phi_n\rangle)$. In the dilute-gas approximation, summing over all these instantons and fluctuations around them gives a contribution to the path integral of the form (switching back from $\varphi$ to $\phi$)
\begin{align}S_{inst.}=\int d^4x\sqrt{-g} \mathcal{P} e^{-S_E}\cos(n\phi),\label{instpot}\end{align}
where $\mathcal{P}$ is a prefactor whose computation whose precise computation has only been achieved in a few cases \cite{Gross:1982cv,Volkov:2000ih}. We will take $\vert \mathcal{P}\vert\sim M_P^4$, since it is at this scale when the effective description breaks down. This is in accordance with previous literature, see \cite{Rudelius:2015xta}.

However, it is also essential to know whether or not $\mathcal{P}$ has an imaginary part. This will be the case if the spectrum of fluctuations around the instanton background has a negative eigenmode. In this case, \eq{instpot} does not have an interpretation as a potential. The instanton is not mediating transitions between different vacua, rather it is telling us that Minkowski space is an unstable configuration which can decay to something else, in the spirit of Witten's bubble of nothing \cite{Witten:1981gj} and of Coleman and DeLuccia's bounce solutions. 

In \cite{Rubakov:1996cn,Rubakov:1996br}, it is argued that the Giddings-Strominger wormhole has exactly one negative eigenmode, and thus $\mathcal{P}$ is purely imaginary. However, as discussed above, the instanton under consideration would be something like \emph{half} a Giddings-Strominger wormhole; we would rotate back to an Euclidean solution after the wormhole throat. It is not clear that the negative mode survives this cutting; however, since we have not computed the instanton prefactor, we do not know wether this solution is truly an instanton or a bounce.

Although the physical effects of an instanton and a bounce are very different, the distinction is not very important if we are concerned only with the possibility of transplanckian axions as candidates for inflation. The instanton would generate a huge potential, cutting down the effective field range to $\phi$'s satisfying
\begin{align} \vert\mathcal{P}\vert e^{-S_E}\cos(n\phi)\sim V_0, \label{constraintinst}\end{align}
where $V_0$ is the required height of the inflaton potential in our model. Similarly, the bounce will initiate a decay to the true vacuum, spoiling inflation and the Universe with it, unless we restrict ourselves to the same field range \eq{constraintinst}. For convenience from now on we will take $\mathcal{P}$ to be real and discuss the physics in terms of axion potentials. The reader should keep in mind the possibility that we are actually talking about bounces. 

\subsection{Consequences for a transplanckian axion}
\label{sec:impli-single}

If we want the full range of the axion $f$ to be available for inflation, we need \eq{instpot} to be very suppressed. In other words, $S_E\gg1$. By looking at \eq{instac1}, this means $M_P n\gg f$. Thus, effects of gravitational instantons constrain the effective axion decay constant. If $f\sim M_P$, the gravitational instantons with low $n$ will not be suppressed. Although strictly speaking we cannot trust our computation for low $n$ since the wormhole throat is of Planckian size, it is reasonable to say $f\sim 10 M_P$ is ruled out. In any case, this kind of effects do rule out parametrically flat axion potentials, within the regime of low energy effective field theory.

The effective field range for the action is determined by the first instanton contribution to become relevant. This is around $S_E\sim1$, or in other words $n\sim \frac{16f}{\sqrt{3\pi}M_P}\approx 5 \frac{f}{M_P}$. This generates a potential of the form $\cos(n\phi)$ which means that in terms of the canonically normalized axion $\zeta=f\phi$, the effective field range is just $\sim M_P/5$. 

This bound is quite similar to those obtained from the WGC, where it was argued that there should be instantons in the theory which creates a potential with higher harmonics which makes the field range subplanckian, for all intends and purposes. The gravitational instantons discussed here are not those demanded by the WGC; the latter are postulated to allow the ``decay'' of the former. However, under some circumstances the two may coincide. 

Axions arising from string theory compactifications typically couple to extended instantonic objects in the theory, for instance euclidean D-branes. The supergravity perspective regards D-branes as extended, extremal black hole solutions \cite{Ortin:2004ms}, and so they are in a sense the stringy version of the instantons discussed above. On the other hand euclidean D-branes are often claimed to be the instantons required by the WGC. We will ellaborate further on this point in section \ref{sec:string}.

The above calculations assumed the validity of Einsteinian gravity up to the Planck scale. This is not generically true in string compactifications, for which this description breaks down at the compactification scale. 
To be more specific, the range of validity of the instanton calculation is bounded by both sides: In order to trust the gravitational instanton solution the wormhole throat should be larger than some scale $\Lambda$ at which extra moduli become light or we start to see stringy physics in some way; at the same time the above computations neglect the contribution of the axion potential coming from the gauge instantons to the stress-energy of the solution. As most of the contribution to the action comes from the vicinity of the throat, we must demand that it is larger than $V_0^{-1}$, where $V_0$ is the scale of the inflationary potential. Thus, we have the bounds
\begin{align}\Lambda^{-1}<a^{1/4}<V_0^{-1}\quad\Rightarrow\quad \Lambda^{-1}<\frac{\sqrt{n}}{(3\pi^3)^\frac14\sqrt{M_Pf}}<V_0^{-1}.\end{align}

\section{Gravitational instantons for multiple axions\label{multiple}}

These bounds on transplanckian decay constants coming from purely quantum gravitational arguments agree with the difficulties found in string theory to provide a transplanckian axion. In \cite{Banks:2003sx} was shown for several examples that in order to have a transplanckian decay constant for an axion, one is always forced to go beyond the safe and controlled regime of the effective theory. New contributions, usually in the form of higher harmonics to the potential, become relevant at scales $\mathcal{O}(M_P)$. In \cite{Grimm:2014vva} the same question was addresed in the context of F-theory, finding similar harmonics that would forbid parametrically large decay constants. People have tried to evade these difficulties by considering models of multiple axions in which you might hope to engineer a direction in the moduli space with an enhanced effective field range. However there is not a completely succesful embedding of any of these models yet in string theory (see \cite{Grimm:2007hs,Ben-Dayan:2014lca,Long:2014dta,Shiu:2015uva,Kappl:2015pxa,Shiu:2015xda} for recent attempts). While the failure to get a single transplankian axion in string theory is quite widely accepted, the same question when there is more than one axion is not clear at all. Here we generalize the computation of the previous section to the case of multiple axions and study the gravitational constrainsts that our instanton imposes over the effective field range available for inflation. We will see that while parametrically flat directions are inconsistent with the presence of gravitational instantons, we can still have a certain numerical enhancement proportional to $\sqrt{N}$ with $N$ being the number of axions, which survives to the gravitational effects. This is in apparent contradiction to the generalized Weak Gravity Conjecture \cite{ArkaniHamed:2006dz,Cheung:2014vva,Rudelius:2014wla,Rudelius:2015xta}. We will clarify this issue in section \ref{twomore} showing a loophole in the conjecture, similar to the example discussed in section \ref{depespec}.

\subsection{Axion-driven multiple field inflation}
\label{sec:align}

Let us consider an effective field theory containing N axions $\phi_i$. The shift symmetries will be broken by non-perturbative effects inducing a scalar potential which is still invariant under discrete shifts $\phi_i\rightarrow \phi_i+2\pi$. The periodicities of the axions define a lattice in the field space of side length $2\pi$. The effective lagrangian of the system takes the form
\begin{align}
\mathcal{L}=\frac12 (\partial_\mu\vec{\phi})\mathcal{G}(\partial^\nu\vec{\phi})- \sum_{i=1}^N\Lambda^4_i(1-\cos(\phi_i))
\end{align}
where in general the kinetic term is not neccessarily canonical. We will refer to this basis as the \emph{lattice basis}. We can go now to the \emph{physical} or \emph{kinetic basis} in which the kinetic terms are canonically normalized by making a rotation to diagonalize the kinetic term
\begin{align}\mathcal{G}=\mathbf{R}^T diag(f_i^2)\mathbf{R}\ ,\quad \vec{\zeta}=\mathbf{R}\vec{\phi}
\label{R}
\end{align}
followed by a field redefinition given by $\hat{\zeta_i}=f_i\zeta_i$. The effective lagrangian in the kinetic basis reads
\begin{align}
\mathcal{L}=\frac12 (\partial_\mu\hat{\zeta_i})^2- \sum_{i=1}^N\Lambda^4_i(1-\cos(R_{ji}\hat{\zeta_i}/f_i))\ .
\end{align}
If we diagonalize now the potential we get a new lattice which depends non-trivially on the kinetic eigenvalues $f_i$ and the rotation angles. The maximum physical displacement in field space (in the absence of monodromy) is given by half of the diagonal of this new lattice.
If $\Lambda_i=\Lambda$ for all $i$, this theory gives rise to N fields with mases $m^2_i=\Lambda^4/f_i^2$ and it turns out that the lightest field is also the one with the largest available field range. In general, in order to obtain succesful inflation along the largest direction in field space we need to tune the parameters $\Lambda_i$. This can lead to some difficulties when trying to embed the model in string theory. However, since this issue is very model dependent, we will ignore it and assume that the parameters $\Lambda_i$ can always be tuned such that inflation would occur along the direction of interest. Our focus will be in proving if that flat direction survives or it is shortened upon including the effect of gravitational instantons.

\begin{figure}[h!]
%\vspace{-10pt}
\begin{center}
\includegraphics[width=0.35\textwidth]{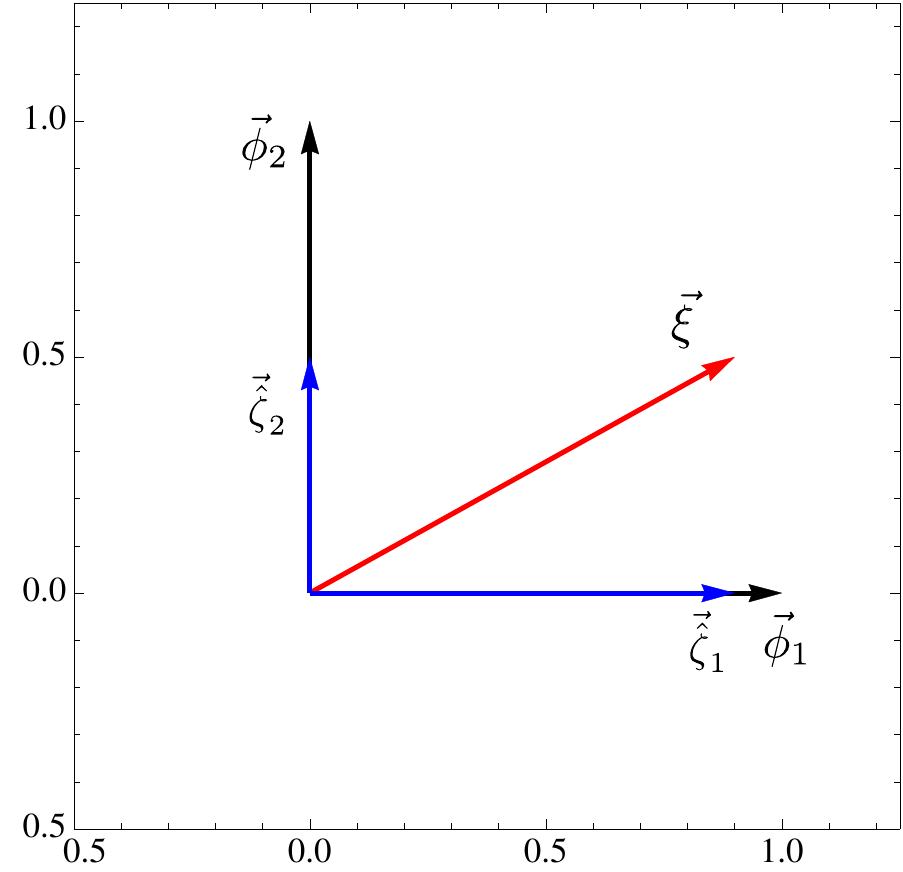}\quad\quad\quad
\includegraphics[width=0.35\textwidth]{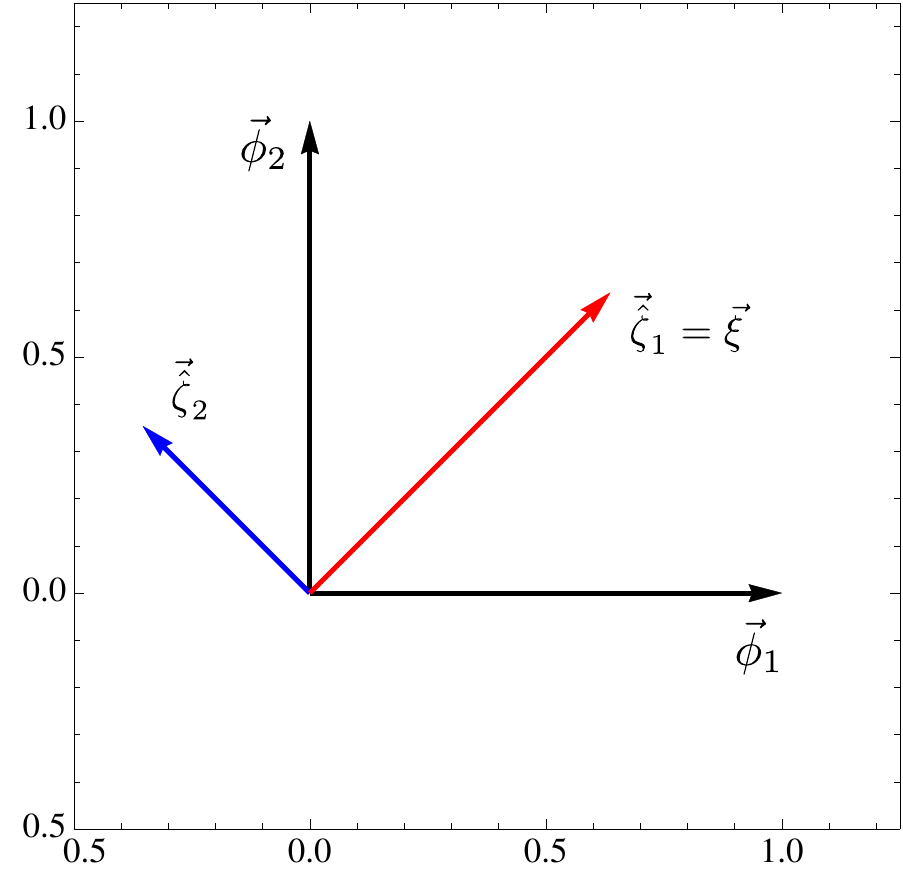}
\end{center}
\vspace{-15pt}
\caption{\footnotesize{Different choices of basis for N-flation (left) and kinetic alignment (right) models.}}
\label{figNaxionslat}
\vspace{-5pt}
\end{figure}

The different ways to achieve large field inflation with multiple axions can be divided in three main proposals:
\begin{itemize}
\item N-flation \cite{Dimopoulos:2005ac}
\item Kinetic alignment \cite{Bachlechner:2014hsa}
\item Lattice alignment \cite{Kim:2004rp}
\end{itemize}
In \emph{N-flation} models the kinetic metric is assumed to be diagonal already in the lattice basis. Then both lattice and kinetic bases are proportional to each other and related by the field redefinition $\hat{\zeta_i}=f_i\phi_i$, so the original lattice is mapped to a new  rectangular lattice with side lengths $2\pi f_i$ (see fig.\ref{figNaxionslat} (left)). The maximum displacement is then given by a collective mode travelling along the diagonal,
\beq
\Delta \Phi=\pi \sqrt{f_1^2+\dots+f_N^2}\ .
\eeq
In the favorable case in which all the decay constant are of similar order $f_1\sim\dots\sim f_N$ we get an enhancement of $\sqrt{N}$.

In models based on \emph{kinetic alignment} the kinetic metric is not assumed to be diagonal when written in the lattice basis. In fact, one can used this freedom to choose the eigenvector $\zeta_i$ with the largest eigenvalue $f_i^2$ pointing along a diagonal of the lattice, such that the maximum displacement is then given by $\pi \sqrt{N} f_i$. This relaxes the requirements to get large field inflation with N axions, since it would be enough to have one entry of the kinetic matrix large, while the others could remain small. For concretenes let us consider two axions. The kinetic basis is given by
\beq
\vec{\hat\zeta}=\left(\begin{array}{cc}f_1&0\\0&f_2\end{array}\right)\mathbf{R}\vec{\phi}=\left(\begin{array}{cc}f_1(\phi_1 \cos\alpha+\phi_2 \sin\alpha)\\f_2(\phi_2 \cos\alpha-\phi_1 \sin\alpha)\end{array}\right)
\eeq
where $f_1,f_2$ are the eigenvalues of the kinetic matrix and we have parametrized the rotation matrix of \eqref{R} as
\beq
\mathbf{R}=\left(\begin{array}{cc}\cos\alpha&\sin\alpha\\-\sin\alpha&\cos\alpha\end{array}\right)
\eeq
It can be easily checked that the maximum physical displacement can be achieved if the eigenvector $\zeta_1$ (assuming $f_1>f_2$) points in the direction of the diagonal of the new lattice, ie. $\alpha=\pi /4$, obtaining 
\beq
\Delta \Phi=\pi \sqrt{2} f_1\ .
\label{displ}
\eeq
%The eigenvectors $\hat \zeta_i$ define a physical lattice with length sides given by $f_i$ and rotate it with respect to the original fundamental lattice.
The situation of perfect kinetic alignment is depicted in fig.\ref{figNaxionslat} (right), in which the inflationary trajectory (red arrow) coincides with the direction determined by $\hat\zeta_1$. For later use the metric in the lattice basis can be written in general as
\beq
\mathcal{G}=\left(\begin{array}{cc}f_1^2\cos^2\alpha+f_2^2\sin^2\alpha&(f_1^2-f_2^2)\cos\alpha \sin\alpha\\(f_1^2-f_2^2)\cos\alpha \sin\alpha&f_1^2\sin^2\alpha+f_2^2\cos^2\alpha\end{array}\right)
\label{metric2}
\eeq
where $\alpha=\pi/4$ for perfect kinetic alignment and $\alpha=0$ for the original models of N-flation.
%~~~~~~~~~~~~~~~~~~~~~~~~~~ FIGURE ~~~~~~~~~~~~~~~~~~~~~~~~~%

%~~~~~~~~~~~~~~~~~~~~~~~~~~~~~~~~~~~~~~~~~~~~~~~~~~~~~~~~~~~%

The Kim-Nilles-Peloso proposal of \emph{lattice alignment} claims to achieve a parametrically flat direction in the potential starting from subplanckian decay constants. However this large direction, although hidden in some choice of basis for the lattice, is there from the beginning and can be made manifest by going to the kinetic basis. Let us consider the lagrangian proposed in the original paper \cite{Kim:2004rp}
 \begin{align}
\mathcal{L}=\frac12 (\partial_\mu\theta_1)^2+\frac12 (\partial_\mu\theta_2)^2- \Lambda_1^4\left(1-\cos\left(\frac{\theta_1}{f_1}+\frac{\theta_2}{g_1}\right)\right)-\Lambda_2^4\left(1-\cos\left(\frac{\theta_1}{f_2}+\frac{\theta_2}{g_2}\right)\right)
\label{VKNP}
\end{align}
Consider for simplicity $f=f_1=f_2$. Perfect lattice alignment occurs if in addition $g_1=g_2$, leading to a flat direction in the potential corresponding to the linear combination $\xi\propto \frac{\theta_1}{g_1}-\frac{\theta_2}{f}$ which is orthogonal to the one appearing in \eqref{VKNP}. A slight misaligment $g_1-g_2\simeq \epsilon$ gives rise to a nearly flat direction in field space with effective field range parametrized by $f_{eff}=g_2\sqrt{f^2+g_1^2}/\epsilon$. This effective decay constant can be done a priori parametrically large by sending $\epsilon\rightarrow 0$. The problem is that by doing this we are also making the kinetic metric diverge. Let us reformulate the model in terms of the lattice and kinetic basis introduced above. The KNP basis $\theta_i$ is related to the lattice basis as follows
\beq
\vec{\phi}=\mathbf{M}\vec{\theta}\ ,\quad \mathbf{M}=\left(\begin{array}{cc}1/f&1/g_1\\1/f&1/g_2\end{array}\right)
\eeq
%where $\mathbf{M}=diag(f_i)\mathbf{R}$ according to our notation. 
In the lattice basis the kinetic term is not canonical and is given by
\beq
\mathbf{\mathcal{G}}=(\mathbf{M}^T)^{-1}\mathbf{M}^{-1}=\frac{1}{(g_1-g_2)^2}\left(\begin{array}{cc}g_1^2(f^2+g_2^2)^2&-g_1g_2(f^2+g_1g_2)\\-g_1g_2(f^2+g_1g_2)&g_2^2(f^2+g_2^2)^2\end{array}\right)
\label{metricKNP}
\eeq
%~~~~~~~~~~~~~~~~~~~~~~~~~~ FIGURE ~~~~~~~~~~~~~~~~~~~~~~~~~%
\begin{figure}[h!]
%\vspace{-10pt}
\begin{center}
\includegraphics[width=0.34\textwidth]{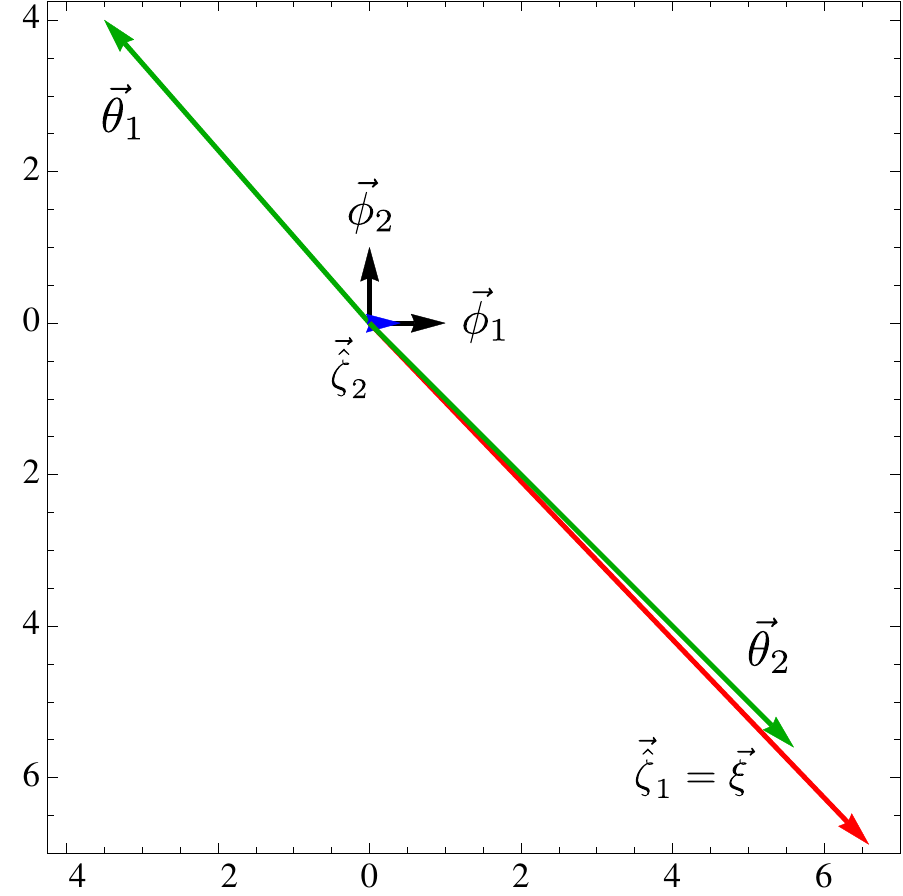}\quad\quad\quad
\includegraphics[width=0.35\textwidth]{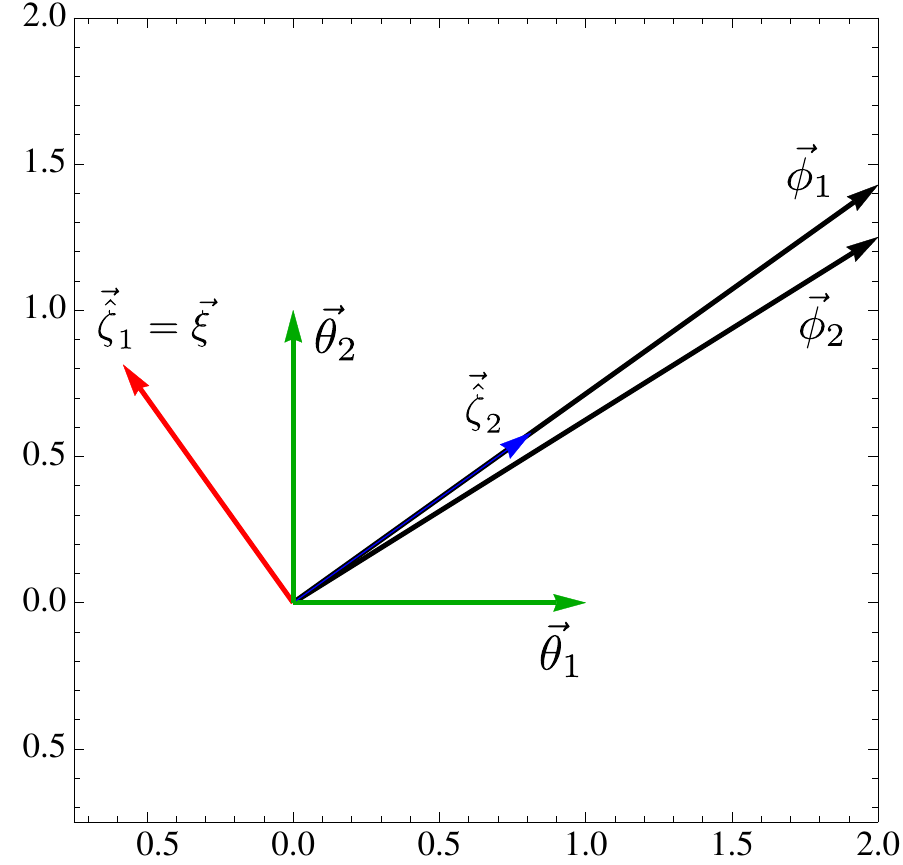}
\end{center}
\vspace{-15pt}
\caption{\footnotesize{Different choices of basis in KNP models. The red line always correspond to the inflationary trajectory.}}
\label{figKNPlat}
\vspace{-5pt}
\end{figure}
%~~~~~~~~~~~~~~~~~~~~~~~~~~~~~~~~~~~~~~~~~~~~~~~~~~~~~~~~~~~%
By diagonalizing $\mathcal{G}$ we can get the kinetic eigenvalues $f_1,f_2$ and the rotation angle $\alpha$ in terms of the initial decay constants $f,g_1$ and $g_2$. Interestingly it turns out that one of the eigenvalues scale as $f_1\sim 1/\epsilon$  while the other as $f_2\sim \epsilon$. Hence an effective transplanckian field range ($\epsilon<1$) is possible only if one of the kinetic eigenvalues is transplanckian. Besides in the limit $\epsilon\rightarrow 0$ we also have that $\text{tan}(\alpha)\rightarrow 1$ so the rotation angle $\alpha\rightarrow \pi/4$ and the system approaches perfect kinetic alignment. In fact, by diagonalizing the metric and doing the field redefinion explained in \eqref{R} (to go to the kinetic basis) it can be checked that the kinetic eigenvector with the largest eigenvalue corresponds indeed to the flat direction that KNP uses for inflation, ie. $\hat \zeta_1\equiv \xi$. Therefore the maximal displacement is still given by \eqref{displ} with the crucial difference that now the eigenvalue $f_1$ is transplanckian. In fig.\ref{figKNPlat} we illustrate the relation between the different choices of basis which play a role in the discussion. In the left figure we show the lattice generated by the potential (basis $\{\phi_1,\phi_2\}$), and the transformation required to go to the KNP basis. In the right figure we take the KNP basis as canonical and show its relation with the other basis. The degeneracy of the matrix $\mathbf{M}$ in the limit of lattice alignment is pictorially translated into the almost degenerate vectors $\{\phi_1,\phi_2\}$.
To recap, a priori we could get a parametrically large flat direction by doing $\epsilon\rightarrow 0$ which is equivalent to increase parametrically the value of the kinetic eigenvalue $f_1$, so the large scale is hidden in the model from the beginning and is not a consequence of a clever change of basis. The remaining question is then if a transplanckian $f_1$ is consistent in a quantum theory of gravity. 

In the following we turn to study the effects of the gravitational instantons over these multiple field inflationary models.

\subsection{Gravitational effects}
\label{sec:impli-multi}

Let us generalize the computation performed in section \ref{single} for the case of N axions with generic non-canonical kinetic terms. The gauge potential sets the periodicities of the axionic fields, but has no relevant effect in the resolution of the Einstein's gravitational equations as was discussed at the end of section \ref{sec:impli-single}. Hence we neglect it from now on as was done in the single field case. The stress-energy tensor of the axion system is now
\begin{align}
T_{\mu\nu}=\partial_\mu\vec{\phi}\mathcal{G}\partial_\nu\vec{\phi}-\frac12(\partial_\mu\vec{\phi}\mathcal{G}\partial^\mu\vec{\phi})g_{\mu\nu} \end{align}
with nonvanishing components
\begin{align}
T_{rr}=\partial_r\vec{\phi}\mathcal{G}\partial_r\vec{\phi},\quad T_{ii}=-\frac{1}{2f(r)}\partial_r\vec{\phi}\mathcal{G}\partial_r\vec{\phi}g_{ii},\ \ i=\psi,\theta,\phi.
\end{align}
Recall that the fields $\phi_i$ are adimensional and define a lattice of periodicities $2\pi$. Dirac quantization then implies
\beq
\int_{S^3}d\vec{\phi}=\mathcal{G}^{-1}\vec{n}
\eeq
with $\vec{n}$ some vector of integer entries. This implies that at large distances the axions must have have an asymptotic profile of the form
\beq
\vec{\phi}\sim \frac{\mathcal{G}^{-1}\vec{n}}{4\pi^2}\frac{1}{r^2}
\eeq
By solving the Einstein's equations we get
\begin{align}
f(r)=\frac{1}{1-\frac{a}{r^4}},\quad \partial_r\vec{\phi}\,\mathcal{G}\partial_r\vec{\phi}= \frac{\vec{n}^T\mathcal{G}^{-1}\vec{n}}{4\pi^4}\frac{f(r)}{r^6},\quad a\equiv\frac{\vec{n}^T\mathcal{G}^{-1}\vec{n}}{3\pi^3}G
\end{align} 
which reduces to the solution of a single field obtained in section \ref{single} upon setting $\mathcal{G}=f^2$. The action of the gravitational instanton becomes
\begin{align}
S_E=\frac{\sqrt{3\pi}}{16}M_P\sqrt{\vec{n}^T\mathcal{G}^{-1}\vec{n}}
\label{instac2}
\end{align}
and induces a potential for the system of axions given by
\beqa
V=\mathcal{P'}e^{-S_E}\cos(n_i\phi_i)
\label{multi-axion-grav-inst}
\eeqa
Notice that the fundamental periodicities of the potential generated by the gravitational instantons are the same as those of the original lattice. Thus if there is some vector $\vec{n}$ for which the action $S_E\sim 1$, the corresponding gravitational induced potential will spoil inflation. Since we have encoded all the information about the periodicties of the physical lattice in $\mathcal{G}$, the effect of gravitational instantons on the realization of large field inflation reduces to see if the 'tricks' used to get large displacements in multiple field inflation also imply $S_E$ to be small.

\begin{figure}[!ht]
%\vspace{-10pt}
\begin{center}
\includegraphics[width=0.4\textwidth]{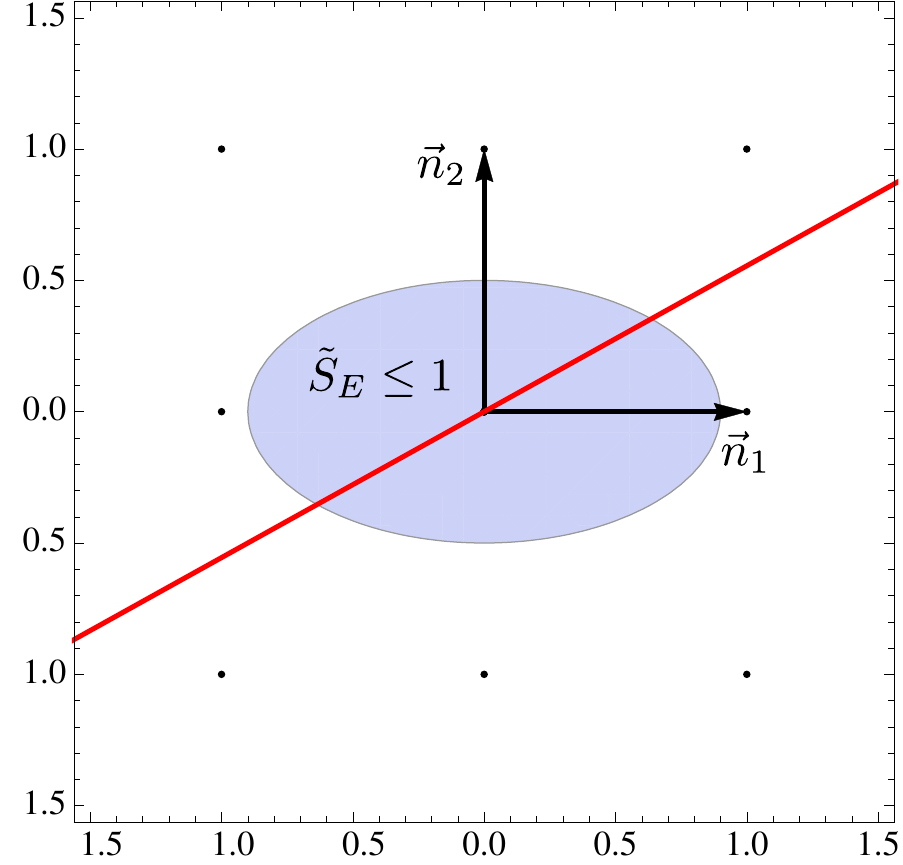}\quad\quad
\includegraphics[width=0.4\textwidth]{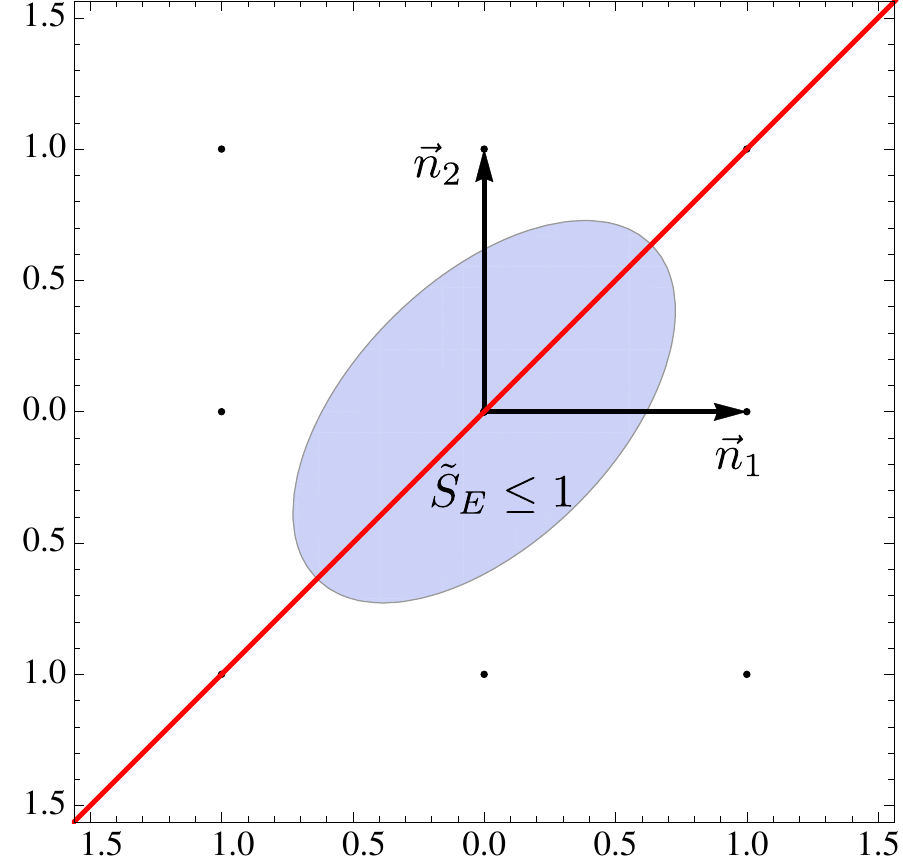}
\end{center}
\vspace{-15pt}
\caption{\footnotesize{Instanton action for a model of N-flation with two axions (left) and a model of kinetic alignment (right). The red line corresponds to the direction of the inflationary trajectory.}}
\label{figNaxions}
\vspace{-5pt}
\end{figure}

The situation of two axions is depicted in Figure \ref{figNaxions} for $\alpha=0$ (left) and $\alpha=\pi/4$ (right), the latter corresponding to the perfect kinetic alignment situation described above. 
%The eigenvectors $\hat \zeta_i$ define a physical lattice with length sides given by $f_i$ and rotate it with respect to the original fundamental lattice. 
The figure illustrates the dual lattice, so for instance the vectors $\vec{n}_1,\vec{n}_2$ correspond to the instantons with integer charges (1,0) and (0,1) respectively. The blue ellipse represents the region for which $\tilde{S}_E=\frac{16}{\sqrt{3\pi}}S_E\leq 1$ and the gravitational instantons induce a non-negligible potential for the axions\footnote{The prefactor $\frac{\sqrt{3\pi}}{16}$ is characteristic of the gravitational instanton under consideration, so it is the same both for one or multiple axions. Hence we plot $\tilde{S}_E$ (the action ommiting $\sqrt{3\pi}/16$) in order to compare directly the constraints for multiple axions with those for a single axion.}. The axis lengths of the ellipse are determined by twice the eigenvalues of the kinetic matrix $\mathcal{G}$ and it is rotated with respect to the fundamental lattice by an angle given by $\alpha$. The red line represents the would be inflationary trajectory in each case. If the physical lattice is such that some point corresponding to a vector $\vec{n}$ lies inside the ellipse, then the gravitational instantons will induce a potential which spoil inflation if the vector $\vec{n}$ is close enough to the inflationary trajectory. This is not the case if the metric eigenvalues remain subplanckian and therefore no point lies within the ellipse, as can be deduced from fig.\ref{figNaxions} for the cases under consideration. But let us take a closer look to understand how this works in general.
%~~~~~~~~~~~~~~~~~~~~~~~~~~ FIGURE ~~~~~~~~~~~~~~~~~~~~~~~~~%
%~~~~~~~~~~~~~~~~~~~~~~~~~~~~~~~~~~~~~~~~~~~~~~~~~~~~~~~~~~~%

In \emph{N-flation} or \emph{kinetic alignment} all the eigenvalues remain subplanckian and the enhancement is generated by travelling along a diagonal of the field space. For simplicity let us consider two axions in the most favorable situation of perfect kinetic alignment, ie. $\alpha=\pi/4$. The kinetic metric is given by \eqref{metric2}, leading to
\beq
\vec{n}^T\mathcal{G}^{-1}\vec{n}=\frac{1}{2f_1^2f_2^2}\left(f_1^2(n_1-n_2)^2+f_2^2(n_1+n_2)^2\right)
\eeq
The maximum displacement takes place along the direction of the eigenvector with the largest eigenvalue. For concreteness, let us consider $f_1>f_2$ so this eigenvector is $\hat \zeta_1= f_1(\phi_1+\phi_2) /\sqrt{2}$, corresponding to a direction with $n\equiv n_1=n_2$ in the lattice basis. Along the inflationary trajectory, the action for the gravitational instanton becomes
\beq
S_E=\frac{\sqrt{3\pi}}{16}M_P\sqrt{\vec{n}^T\mathcal{G}^{-1}\vec{n}}=\frac{\sqrt{3\pi}}{16}M_P\frac{\sqrt{2}n}{f_1}
\eeq
Notice that this action is a factor $\sqrt{2}$ bigger comparing to the one obtained in the single field case \eqref{instac1}, while the maximum displacement \eqref{displ} is a factor $\sqrt{2}$ larger. This result can be generalize to N axions obtaining that the same factor $\sqrt{N}$ which enhances the effective field range, also appears in the instanton action suppressing the amplitude of the induced gravitational potential.

The same results apply for N-flation, where the only difference is that both kinetic and lattice basis are proportional so $\alpha=0$ and the decay constant entering in $S_E$ is not the largest one but the Pythagorean sum of all of them.

\begin{figure}[!ht]
%\vspace{-10pt}
\begin{center}
\includegraphics[width=0.4\textwidth]{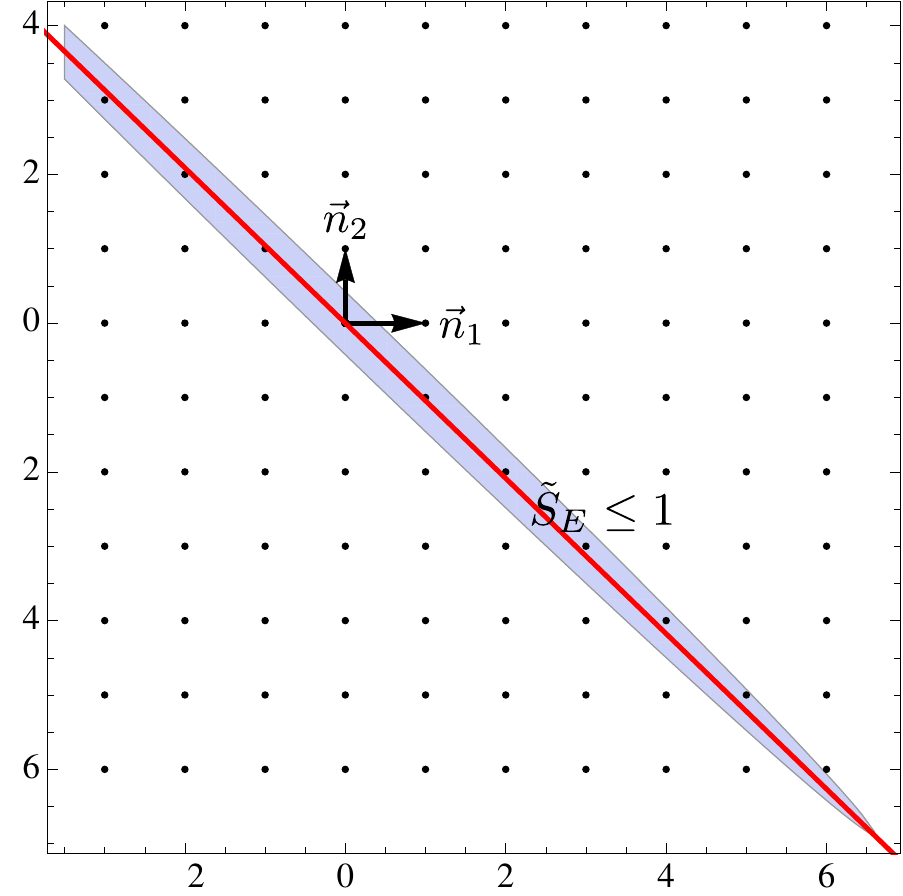}
\end{center}
\vspace{-15pt}
\caption{\footnotesize{Gravitational effects for a model of lattice alignment. The blue ellipse represents the region for which the instanton action is small and its effect over the axion can not be neglected. The red line corresponds to the direction of the inflationary trajectory.}}
\label{figknp}
\vspace{-5pt}
\end{figure}

The case of \emph{lattice alignment} is however different. We have seen that a large effective decay constant in the proposal of KNP implies a parametrically large eigenvalue of the kinetic matrix $\mathcal{G}$. Hence the inverse of the metric has a nearly zero eigenvalue and the effects of the gravitational instantons are huge. In more detail, along the kernel direction the action will scale as $S_E\sim \epsilon^2$ being $\epsilon$ the misalignment parameter, so if $\epsilon<1$ (in order to have a transplanckian effective field range) then the action is automatically $S_E\lesssim 1$. It can be checked that the kernel direction is indeed the nearly flat direction used for inflation, so there is no way out. A vector $\vec{n}$, no matter how big its components, will have nearly vanishing action as long as it is close enough to the kernel direction and will spoil inflation. The result becomes even more clear by looking at Figure.\ref{figknp}. In the limit of lattice alignment one of the kinetic eigenvalues tends to zero while the other diverges. This implies that the ellipse is very thin in one axis and very long in the other one, so it captures most of the points with $n_1=-n_2$ (all of them if there is no misalignment). This implies that the instantons with charges $n_1=-n_2$ will have a nearly zero action and will induce a huge potential for the axions. In addition, the inflationary trajectory is identified with the direction of the eigenvector with the largest eigenvalue, implying that these instantons indeed generate a potential along the inflationary trajectory and spoil inflation. Each of the points lying within the ellipse can be seen as a higher harmonic contribution which reduces dramatically the field range. One could hope to engineer a concrete model in which all these contributions are suppressed enough to get succesful inflation, but then the advantage of invoking an axion in order to have controll over higher order corrections is partially lost. One should really keep trace of all these non-negligible non-perturbative effects, which is clearly not an easy task. In this sense a model based in lattice alignment does not provide any improvement with respect to a model of a single axion.
%~~~~~~~~~~~~~~~~~~~~~~~~~~ FIGURE ~~~~~~~~~~~~~~~~~~~~~~~~~%
%~~~~~~~~~~~~~~~~~~~~~~~~~~~~~~~~~~~~~~~~~~~~~~~~~~~~~~~~~~~%

Our results indicate that having multiple axions might help to relax the gravitational constraints found for the effective field range of a single axion, as long as there are not intrinsic transplanckian scales hidden by any change of basis. To check that, one can always go to the lattice basis such that all the information about the length scales is encoded in the kinetic matrix, and check that the eigenvalues remain subplanckian. If this is the case one can still have an enhancement of the decay constant by travelling along a diagonal of the field space. However this enhancement is given at most by a factor of $\sqrt{N}$, so for N fixed one can never get a parametrically large field range. Even so, one could think that this result is in contradiction with the results of section \ref{single}, since one could always integrate out all massive degrees of freedom, except the one corresponding to the inflationary trajectory, ending up again with a single inflationary model with a transplanckian field range but this time without a gravitational instanton to constrain it. Therefore we must be missing some information when integrating out naively the massive degrees of freedom. The solution to this 'paradox' is clarified in section \ref{twomore}, but before that, let us comment about two assumptions we have made in our computation.

First, we have assumed that the fundamental periodicities of the axions are the ones determined by the scalar potential. However, if $V\propto \cos(\phi/f)$, the potential is also periodic under shifts $\Delta\phi=2\pi mf$ where $m$ can be any positive integer. This is indeed what happens in most of the attempts to embed alignment inflation in string theory, where the induced potential is $V\propto \cos(m\phi/f')$ with $f'=mf$ and $m>1$. Thus a natural question would be what changes in the study of the gravitational instantons if $m>1$. The answer is that all of our results still apply in the same way, but we are missing in addition some gravitational instantons which could modify the profile of the potential but not reduce the field range. The generalization of the potential induced by the instantons if $m>1$ reads
\beq
V\sim e^{-S_E}\cos\left(\frac{n\phi\, m}{f'}\right)\ ,\quad S_E\sim \frac{nM_P\, m}{f'}
\label{sublattice}
\eeq 
The instanton to which we were assigning the unit charge $n=1$ actually would correspond to that of charge $n'=m$. It is clear from \eqref{sublattice} that by redefining $f'=mf$ we are still capturing the effects of all instantons with $n'\geq m$, but we are missing the information about instantons with $n'<m$. However these latter instantons induce a potential periodic only under bigger shifts $\Delta\phi>2\pi f$. Hence although they might modulate the inflationary potential in a non negligible way if the instanton action is small enough, they are not relevant for our discussion since they do not reduce the effective field range. This can be easily generalized to more than one axion. The periodicities set by the potential generate a lattice $L$, but they are also compatible with any sublattice $S\subset L$. Since the dual lattice $L^*\subset S^*$ is contained in the dual of $S$, $S^*$, by considering $L$ we are missing some instantons present in $S^*$ but not in $L^*$, which do not respect the periodicity of the potential but instead have a bigger period.

Finally, here we have assumed $M_P$ to be a constant. However, the Planck mass will also receive quantum corrections from diagrams involving the $N$ axions running in the internal loops, so $M_P$ will grow as $\sqrt{N}$, cancelling the Pythagorean enhancement from travelling along the diagonal \cite{Dimopoulos:2005ac,Grimm:2007hs}. It has been argued that suitable cancellations can occur such that even if a parametrically large field range is not possible, one might still get a sufficient large field range to provide inflation. Studies regarding the diameters of axion moduli spaces in Calabi-Yau manifolds have been carried out in \cite{Rudelius:2014wla}, showing the difficulties arising to have an enhancement of a moduli space diameter while keeping the overall volume small (see also \cite{Bachlechner:2014gfa} for a more optimistic view). We just want to remark that the gravitational effects which are usually argued to be behind the difficulties from getting transplanckian field ranges in string theory might actually be suppressed for the case of multiple axions. Some of the gravitational instantons which are assumed to be present in the IR, might actually not exist due to the selection rules of some $Z_N$ discrete gauge symmetries in the UV, as we proceed to explain in the following.

\section{Two is more than one}\label{twomore}

We now want to point out what seems to be a paradox between the results of sections \ref{single} and \ref{multiple}. We will consider a system with two axions $\vec{\phi}$. In the basis in which the lattice is diagonal, we will have a nontrivial kinetic term $\mathcal{G}$. The lagrangian will be
\begin{align}\mathcal{L}=\frac12 (\partial_\mu\vec{\phi})\mathcal{G}(\partial^\nu\vec{\phi})- \sum_{i=1}^N\Lambda^4_i(1-\cos(\phi_i)).\label{sec5lag}\end{align}
We will couple this system minimally to gravity, and also assume $\Lambda_i\ll M_P$. As discussed above, kinetic alignment provides successful inflation. In other words, we take $\Lambda_i=\Lambda$ and 
\begin{align}\mathcal{G}=\mathbf{R}^T \left(\begin{array}{cc}f^2&0\\0&g^2\end{array}\right)\mathbf{R},\end{align}
where $\mathbf{R}$ is the rotation matrix which takes the vector $(1,0)$ to the diagonal $\frac{1}{\sqrt{2}}(1,1)$. Indeed, if one defines $\vec{\zeta}=\mathbf{R}\vec{\phi}$, then the potential becomes
\begin{align}\Lambda^4(2-\sum_{j=1}^2\cos(2\pi R_{ji}\zeta_i)=2\Lambda^4\left[1-\cos(\frac{\zeta_1}{\sqrt{2}})\cos(\frac{\zeta_2}{\sqrt{2}})\right].\label{lagpuz}\end{align}
By changing to $\hat{\zeta_1}=\sqrt{2}f\zeta_1$, $\hat{\zeta_2}=\sqrt{2}g\zeta_2$, we see that this field theory describes $2$ fields with masses $\Lambda^4/g^2$ and $\Lambda^4/f^2$. If $g\ll f$, we may take the latter as an inflaton. The potential along its direction has a periodicity $\sqrt{2} f$.

So far, we have only described again the kinetic alignment mechanism of \cite{Bachlechner:2014hsa} discussed in the previous section. However, we may now integrate out all degrees of freedom with mass $\Lambda^4/g^2$ and higher, as long as we are away of $\zeta_1\sim \pi/\sqrt{2}$, which is where inflation ends. We will be left with a theory of a single transplanckian axion which would seem to have an enhanced periodicity $\sqrt{2} f$.

From the results of section \ref{single}, we know that such a theory has gravitational instantons contributing potentials of the form $\cos\left(n\frac{\hat{\zeta}_1}{\sqrt{2}f}\right)$, with $n$ any integer. As discussed there, these gravitational instantons reduce the effective field range, constraining $f< M_P$ and thereby spoiling inflation. 

 However, the gravitational instantons contributing to the action can also be studied before integrating out the heavy axion, and we know from the results of section \ref{multiple} that in this case inflation with a transplanckian field range is indeed possible. What is going on?
 
 A look at the instanton action \eq{instac2} tells us that any instanton not in the direction of $(1,1)$ will pick up a contribution proportional to $1/g$ and hence will be tremendously suppresed. When computing instanton corrections to the path integral, a consistent criterion is to pick all the configurations with action smaller than some  preestablished value; in this section we will drop any instantons whose action goes as $1/g$, in accordance with our initial statement that $f\gg g$. Notice however that configurations with actions of order $M_P/f$ will be large (since $f$ is subplanckian) but will be retained nevertheless; otherwise, we would not be able to analyze the potential generated by the gravitational instantons. Also, $M_P/f\gg 1$ is the regime when the instanton throat, etc. are larger than the Planck length and so our effective field theory computations can be trusted.

Let us compute the action of an instanton with $\vec{n}=n(1,1)$:
\begin{align}S_E^{\text{instanton}}(n)=\frac{\sqrt{3\pi}}{16}\frac{n\sqrt{2} M_P}{f}=\frac{\sqrt{3\pi}}{16}\frac{M_P}{\sqrt{2}f}(2n).\end{align}
The instanton generates a potential of the form $\cos\left(2n \frac{\hat{\zeta_1}}{\sqrt{2} f}\right)$. That is, gravitational instantons only generate even harmonics in terms of the periodicity of the effective transplanckian axion established by the gauge potential.

Clearly, the effective field theory of one axion is missing some constraint coming from the UV. A selection rule forbidding odd-order harmonics is highly reminiscent of a discrete symmetry \cite{Banks:2010zn,BerasaluceGonzalez:2011wy, Ibanez:2012wg, BerasaluceGonzalez:2012vb}. We will now study the precise way in which the IR theory forbids the presence of these odd charge instantons: it can do so because in the $g\rightarrow 0$ limit there is another set of instantons of low action which contribute the path integral, namely, charged axionic strings.

\subsection{The string}
\label{sec:charged-string}

Gravity not only enables us to build an instanton for any axion we may have, but it also allows us to build the dual object: a string around which the axion has the asymptotic profile $\vec{\phi}=\theta\vec{m}$, where $\theta$ is the angle around the string. The instanton couples electrically to the axion field, whereas the string has a magnetic coupling of the kind
\begin{align}\int_{\text{string}} d\vec{\phi}.\end{align}

So, we are led to study solutions of the coupled Einstein-axion system with cylindrical symmetry and monodromy $2\pi \vec{m}$ as we circle around the string. Solutions to this system have been considered before in the literature \cite{Kaloper:1992kw,PhysRevD.33.333,Harari:1989fb,Gregory:1988xc}. Asymptotically, the stress energy tensor in cylindrical coordinates is of the form
\begin{align}T_\mu^\nu=\frac{C}{r^2}\text{diag}(1,-1,1,1).\end{align}
 It turns out \cite{Harari:1989fb} that of the solutions of Minkowskian Einstein's equations corresponding to the above mondromy only one has locally flat asymptotics, and it corresponds to a very unphysical configuration in which there is an event horizon outside which the metric is not static. Physically, what happens is that the constant stress-energy implied by an asymptotic profile with the above monodomy acts as a fluid with larger and larger stresses as we go away from the string, which in turn curves the space at asymptotic regions. To compensate and get a flat solution one must put large negative sources of stress-energy at the core of the string, and these cause the losss of staticity of the solution.

Although it is not possible to have an asymptotically flat Minkowskian metric describing one such string, it may be possible to have one describing \emph{two} strings with opposite monodromies, so that asymptotically the stress-energy vanishes. This suggests (and this is indeed the case) that although there are no asymptotically flat Minkowskian solutions, there might be asymptotically flat \emph{Euclidean} ones. Physically, these would describe the nucleation of a string- anti string pair, or one could have closed strings as loop diagrams contributing to vacuum to vacuum amplitudes. The crucial element ensuring the existence of well-behaved asymptotically flat Euclidean solutions when there are no Minkowskian ones is the swap of the sign of the kinetic term of an axion when continued to the euclidean \cite{ArkaniHamed:2007js}. This was also crucial to ensure the existence of our instantons above.

We thus propose to consider a metric tensor for the string of the form (analytic continuation of \cite{Harari:1989fb,Kaloper:1992kw}, whith ``wrong'' kinetic term sign)
\begin{align}ds^2=\left(1+\frac{1}{\sigma}\log\left(\frac{r}{r_0}\right)\right)dt^2+dz^2+\left(1+\frac{1}{\sigma}\log\left(\frac{r}{r_0}\right)\right)^{-1}dr^2+\mu\sigma r^2d\theta^2\label{str0}\end{align}
The axion string must satisfy $\int_{S^1} d\vec{\phi}=2\pi\vec{m}$, where the $S^1$ is any $S^1$ which encloses the string. This immediately implies by symmetry that $\partial_\theta \vec{\phi}=\vec{m}$, or $\vec{\phi}=\vec{m}\theta$. Also, $\mu=\vec{m}^T \mathcal{G}\vec{m}$. Notice the branch cut for the axion; this creates a ``Dirac domain wall'', similar to the Dirac string in the singular gauge description of monopoles. Instantons crossing this wall should contribute a trivial phase. The parameter $r_0$ can be related to the total length of the Dirac string in some contrived way; we will leave it and study the results as a function of $r_0$. 

\begin{figure}\begin{center}\includegraphics[scale=1.2]{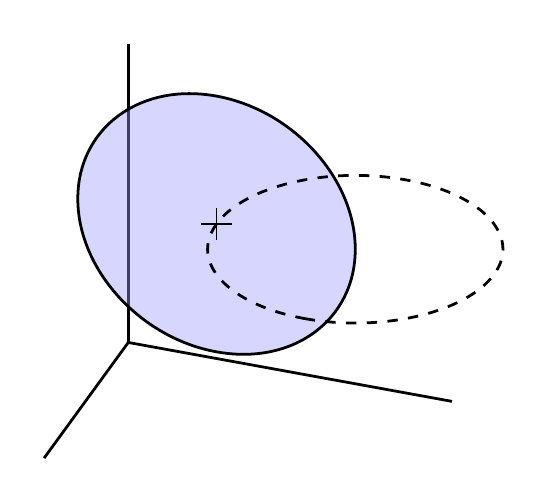}\end{center}
\caption{\footnotesize Pictorial representation of the euclidean string (thick black line) and its associated Dirac domain wall (blue shaded circle). When we move an instanton around the dashed circle, it picks an extra phase as it crosses the domain wall. This phase must be trivial for the theory to be consistent.}
\label{domwall}
\end{figure}

This metric will describe the near-field limit of a closed string. The Dirac domain wall will therefore be a disc, whose boundary is the actual string (see figure \ref{domwall}).

The metric \eq{str0} has a coordinate singularity at $r_h=r_0\exp(-\sigma)$, but as usual when continuating to the Euclidean the space is much nicer. We must take the time coordinate as periodic in order to avoid a conical singularity \cite{Wald}. The period of the imaginary time coordinate is $4\pi r_h$, i.e. $2\pi$ times twice the horizon radius, as usual. 

On the other hand, the metric we have built for the euclidean string is not asymptotically flat. This is not a problem, as it describes the fields near the core of the string; one expects that an asymptotically flat solution exists for a closed string, and that the above metric is a well approximation near the core of the string. However, this means that for realistic solutions one should impose an upper cuttof $R$ after which the approximation \eq{str0} is no longer good; typically, $R$ will be of order of the radius of the closed string. 

We have 
\begin{align}\frac12\sqrt{g} (\partial_\mu \vec{\phi})\mathcal{G}(\partial^\mu\vec{\phi})=\frac{\mu}{\sqrt{g_{\theta\theta}}}= \frac{\sqrt{\mu}}{2\sqrt{\sigma}r}.\end{align}
And thus the action per unit length is
\begin{align}\frac{S_E^{\text{string}}}{2\pi R}\approx8\pi^2r_h\int_{2r_h}^R  \frac{\sqrt{\mu}dr}{\sqrt{\sigma}r}=\frac{8\pi^2 r_h\sqrt{\mu}}{\sqrt{\sigma}}\log\left(\frac{R}{2r_h}\right)=\frac{8\pi^2 r_h\sqrt{\mu}}{\sqrt{\sigma}}\left[\frac{1}{\sigma}+\log\left(\frac{R}{2r_0}\right)\right].\end{align}

The point here is that, as a function of $\sigma$, the action of the string can be made arbitrarily small for $\sigma\rightarrow\infty$, which corresponds to vanishing small mass per unit length. So in principle the action of the above instanton can be made arbitrarilly small. A cutoff however is imposed by the fact that $r_h$ should be greater than the Planck length, otherwise the above solution should not be trusted. Since $r_0$ will be of order $R$, this imposes $\sigma< \log(r_0/l_p)$ so we arrive at
\begin{align}\frac{S_E^{\text{string}}}{2\pi R}>8\pi^2\sqrt{\mu}\frac{1}{\sqrt{\log\left(\frac{r_0}{l_p}\right)}}\left[\frac{1}{\log\left(\frac{r_0}{l_p}\right)}+\log\left(\frac{R}{2r_0}\right)\right]\end{align}
The good thing about the logarithms is that they contribute things of order 1, 2 at most. Also, the radius of the string must be several orders of magnitude larger than the Planck length (i.e, $R\gtrsim l_p$), so we end up with a bound (restoring the Planck mass)
\begin{align}S_E^{\text{string}}\gtrsim 16\pi^3\sqrt{\frac{\vec{m}^T \mathcal{G}\vec{m}}{M^2_p}}.\label{stract}\end{align}
%Notice that the above expression implies that for $g\ll f$ the instanton $\vec{m}=(1,0,\ldots)$ has an action which scales like 

\subsection{Putting everything together}
\label{sec:putting}

Now that we have the action of the Euclidean strings, we have to ask ourselves under which circumstances do they contribute signficantly to the path integral. Recall, as discussed above, that we will neglect contributions to the path integral of action $\sim M_P/g$ but not those which depend on $f$. 

With this criterion, instantons with $\vec{n}=(n_1-n_2)$ with $n_1+n_2$ have an action which scales as $M_P/g$ and are therefore suppressed. However, their dual strings (say, the string with $\vec{m}=(1,0)$) are not suppressed; from \eq{stract} their action goes as
\begin{align}S_E^{\text{string}}\gtrsim 16\pi^3 M_P\sqrt{\frac{f^2(m_1+m_2)^2+g^2(m_1-m_2)^2}{2}}.\end{align}
Thus, when we compute the IR effective theory for $\hat{\zeta}_1$, it is not consistent  to integrate out $\hat{\zeta}_2$ and  retain the gravitational instantons which generate the even-harmonics potential for  $\hat{\zeta}_1$. If we want the latter, we must allow for any other configuration which goes as $f/M_P$ in the path integral, such as the strings. The effective theory in the IR is not a single axion of periodicity $\sqrt{2} f$ coupled to gravity; it also contains euclidean strings which change the quantum theory. 

These strings constrain the allowed gravitational instantons, forbidding those of odd charge.  As discussed above, the strings generate a Dirac domain wall. Consider a configuration with a string with vector $\vec{m}$ and an instanton with $\vec{n}$. If we move the instanton so as to cross the Dirac domain wall, the phase should be (a multiple of) $2\pi$. Actually, the phase the instanton gets upon moving along an $S^1$ with linking number 1 with the string is
\begin{align}\int_{S^1} \vec{n}\cdot d\vec{\phi}=2\pi\, \vec{n}\cdot\vec{m}.\end{align}
To write this amplitude in terms of IR quantities, we must substitute $\vec{n}=\frac{n}{2}(1,1)$. $n$ could be any integer and still respect the periodicity of the lattice. However, for $\vec{m}=(1,0)$ that the above phase becomes  just $\pi n$. Consistency of the path integral thus requires that only even-order harmonics contribute. 

From the UV point of view, what is happening is clear. The instanton discussed in the previous paragraph would lift to an instanton with $\vec{n}=\frac{1}{2}(1,1)$. This is not a vector of the instanton lattice, and hence the instanton is not allowed. However, the $\IZ_2$ discrete symmetry operating in the IR is harder to see. Notice that the absence of odd harmonics in the potential is only strictly true in the limit $g\rightarrow 0$; otherwise we have gravitational instantons breaking the $\IZ_2$ symmetry, even if they are very suppressed.

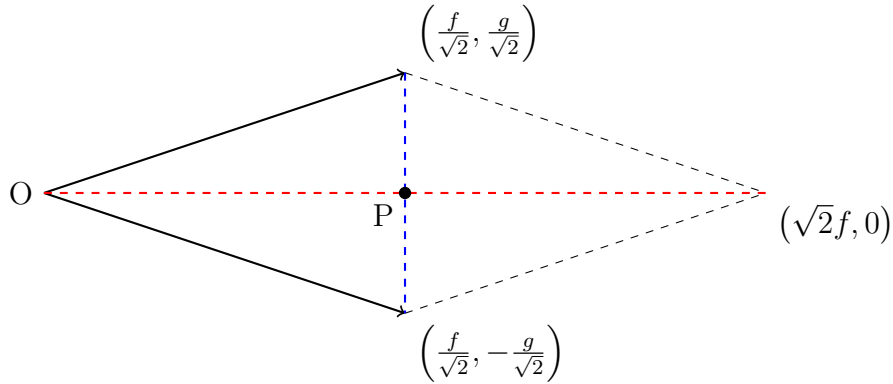
\begin{figure}[!htp]
\begin{center}\begin{tikzpicture}[scale=1.6]
\draw[thick,->] (-3,0) -- (0,1) node [anchor= south west]{$\left(\frac{f}{\sqrt{2}},\frac{g}{\sqrt{2}}\right)$};
\draw[thick,->] (-3,0) -- (0,-1) node [anchor= north west]{$\left(\frac{f}{\sqrt{2}},-\frac{g}{\sqrt{2}}\right)$};
\draw[dashed] (0,1) -- (3,0);
\draw[dashed] (0,-1) -- (3,0);
\draw[thick,dashed,red] (-3,0) -- (3,0);
\draw[thick,dashed,blue] (0,1) -- (0,-1);
\draw (0,0) node[shape=circle,draw,fill=black,inner sep=1.5pt]{};%{$\left(\frac{f}{\sqrt{2}},0\right)$};
\draw (0,0) node[anchor=north east,]{P};
\draw (-3,0) node[anchor=east,]{O};
\draw (3,0) node[anchor=north west ]{$\left(\sqrt{2}f,0\right)$};
\end{tikzpicture}
\end{center}
\caption{\footnotesize Basis of the axion lattice with canonically normalized axions. The diameter of the axion lattice is given by the red dashed line. As $g\rightarrow 0$, the two elements of the lattice basis move along the blue dashed line to degenerate to the segment $\overline{OP}$, resulting in a one-dimensional lattice with half the periodicity.}
\label{deglattice}
\end{figure}

In the limit $g\rightarrow 0$, the two antidiagonal corners of the axion lattice are at zero distance from each other. The lattice in which the canonically normalized fields $\vec{\phi}$ live, which is spanned by the vectors $\frac{1}{\sqrt{2}}(f,g)$ and $\frac{1}{\sqrt{2}}(f,-g)$ degenerates to a single straight line generated by $\frac{1}{\sqrt{2}}(f,0)$ (see figure \ref{deglattice}). As a result, an extra lattice point appears on the diagonal of the unit cell. So although it indeed seems that the axion lattice has a diagonal of length $2\pi\sqrt{2} f$, in the degenerate limit the extra lattice point cuts that periodicity by half. As a result, only even-order harmonics are allowed in the potential. In other words, a theory with two axions in which one becomes very massive cannot be reduced to the effective theory of the remaining axion, but rather requires the inclusion of something else, an additional $\IZ_2$ symmetry: Two is more than one.

If we had actually taken the $g\rightarrow 0$ limit the $\IZ_2$ gauge symmetry would have been exact and even the gauge potential used for inflation would have had to obey it. The theory would contain a single effective axion of subplanckian periodicity $2\pi f/\sqrt{2}$, and that would be the end of the story. By allowing a small, explicit breaking of the symmetry, we can have a transplanckian axion with highly suppressed gravitational contributions, allowing for tranplanckian inflationary regimes. Speaking loosely, one may degenerate the axion lattice in a precise way so that the gravitational effects ``think'' that the effective axion is subplanckian, but the gauge effects know it is not. 

The above discussion generalizes in a straightforward manner to $N$ axions. In this case, there will be a $\IZ_N$ gauge symmetry, which in the UV will come from the degeneration of all the lattice basis vectors to the form $\frac{1}{\sqrt{N}}(f,0,\ldots)$.

\section{Aspects of string theory realizations}
\label{sec:string}

In earlier sections we have considered gravitational instantons in theories of gravity coupled to axions, as part of the effects present in a complete quantum gravitational theory. In this section, we focus on the interplay between our results and expectations in string theory compactifications realizing the axion models described.

\subsection{Generalities}
\label{string-gen}

A first observation is that, in order to produce viable single-field inflation models, the string theory compactifications must deal with the challenge of rendering all fields except the axion(s) massive at a hierarchically large scale, so that they do not interfere with the axion inflationary scenario. This typically requires a relatively large breaking of supersymmetry, in order to get rid of the saxion superpartners. Therefore, in our consideration we pay special attention to often-forgotten features arising in the absence of supersymmetry.

In string theory, axions are ubiquitous. For instance they can arise very generically from the KK compactification of higher-dimensional $p$-form fields over $p$-cycles in the internal space (see e.g \cite{Silverstein:2008sg,McAllister:2008hb,Berg:2009tg,Flauger:2009ab,Palti:2014kza,Marchesano:2014mla,Blumenhagen:2014gta,
McAllister:2014mpa,Franco:2014hsa,Blumenhagen:2014nba} for such axions as inflaton candidates); they also arise from D-brane positions in toroidal compactifications or in large complex structure limits of Calabi-Yau compactifications (see \cite{Hebecker:2014eua,Ibanez:2014kia,Hebecker:2014kva,Ibanez:2014swa} for such axions as inflaton candidates). For concreteness and for their genericity, we focus on axions coming from the RR $p$-form potentials integrated over $p$-cycles.

In general, many of these axions can get masses from fluxes in general compactifications. In applications to inflation, one assumes that this happens at a high scale\footnote{Actually, is is possible to consider models in which some axion receives a hierarchically small contribution to its potential from the fluxes. In those cases however  the structure of the flux potential is such that it induces axion monodromy (e.g. the F-term axion monodromy inflation models \cite{Marchesano:2014mla}, see also \cite{Blumenhagen:2014gta,
McAllister:2014mpa}), for which our analysis does not apply.}, so that one is left with a low energy theory with a reduced set of axions. 

The non-perturbative contributions to the potential for these axions (the gaugino condensate in phenomenological models like natural inflation or its diverse aligned multi-axion variants) are realized in terms of non-perturbative effects on gauge sectors arising from wrapped D-branes in the model. For instance, many models are based on type IIB theory with axions coming from the RR 4-form on 4-cycles, and the 10d type IIB axion. The relevant gaugino condensate non-perturbative effects arise from gauge sectors on D7-branes on 4-cycles (and possibly with worldvolume gauge backgrounds inducing D3-brane charge coupling to the 10d type IIB axion). 

In addition to these field theory non-perturbative effects, there are non-perturbative effects from euclidean D-brane instantons. In general, for any axion in string theory there are such D-brane instanton effects which induce corrections in the 4d effective action \footnote{Actually, for axions made massive by fluxes, the corresponding instantons are absent due to Freed-Witten consistency conditions \cite{KashaniPoor:2005si}; but these are precisely the axions which we have assumed have been integrated out in our effective theories.}. In the above example, of type IIB with axion from the RR 4-form (and the universal axion), the corresponding instantons are euclidean D3-branes wrapped on 4-cycles (possibly magnetized, inducing D(-1)-brane instanton charge, coupling to the universal axion). Actually, the non-perturbative gaugino condensate superpotential can be described in terms of `fractional'\footnote{The fractionalization of instantons in an $SU(N)$ gauge theory is manifest in the F-theory lift, in which the elliptic fibration over $N$ D7-branes pinches off into $N$ component 2-cycles, each of which can be wrapped by an instanton (described as an M5-brane on the 2-cycle of the elliptic fiber fibered over the base 4-cycle).}  D3-brane instantons wrapped on the D7-brane 4-cycle (and with the same worldvolume gauge background). D3-brane instantons which do not have a gauge theory counterpart are dubbed exotic or stringy in the literature \cite{Blumenhagen:2006xt,Ibanez:2006da,Florea:2006si,Blumenhagen:2009qh}.

In supersymmetric backgrounds, the kind of superspace interaction induced by euclidean D-brane instanton effects depends on the structure of exact fermion zero modes (see \cite{GarciaEtxebarria:2008pi} for discussion). For instance, superpotential terms (and therefore contributions to the scalar potential) are generated only by instantons with exactly two fermion zero modes (to saturate the $d^2\theta$ measure) \cite{Witten:1996bn}. Instantons with additional fermion zero modes induce non-perturbative corrections to higher F-terms (which locally in moduli space can be written like D-terms), namely higher-derivative terms or multi-fermion operators \cite{Beasley:2004ys}. However, in the presence of supersymmetry breaking effects, these higher F-terms can descend to contributions to the superpotential (or in non-susy language, to the scalar potential), by contracting the additional external legs with insertions of the supersymmetry breaking operators \cite{Uranga:2008nh}; more microscopically, the susy breaking sources can lift the additional fermion zero modes of the instanton, to allow them to contribute to the scalar potential (see e.g. \cite{Billo:2008sp,Billo:2008pg} and references therein). Therefore, a general lesson in non-supersymmetric situations is that fairly generically all D-brane instantons induce corrections to the scalar potential; the only price to pay is the appearance of suppression factors in $M_{\rm susy}/M_s$, where $M_{\rm susy}$ is the susy breaking scale. Since our main interest lies in applicactions to large-field inflation, in which the inflation scale is high and supersymmetry is broken substantially (e.g. to remove the saxion degrees of freedom), our setup is that all D-brane instantons contribute to the potential of the corresponding axions\footnote{This is similar to the implicit assumption we are taking for gravitational instantons, which in the presence of supersymmetry may have extra fermion zero modes and contribute to higher-derivative terms \cite{Anglin:1992ym}.}. We will often omit the very model-dependent prefactors depending on $M_{\rm susy}/M_s$.

\subsection{D-brane instantons and gravitational instantons}
\label{string-match}

String theory thus provides a set of non-perturbative objects coupling to the low-energy axions, and inducing non-perturbative contributions to the scalar potential of the form
\beqa
V_{\rm D-inst} \sim {\cal P} e^{-S_{\rm D-inst}} \, [\, 1-\cos(\sum_i n_i\phi_i)\, ]
\label{dbrane-inst-pot} 
\eeqa
where $n_i$ are integer charges under the axions $\phi_i$, normalized to $2\pi$ periodicity, ${\cal P}$ is an  instanton dependent prefactor, and $S_{\rm D-inst}$ is the D-brane instanton action.

For instance, in the example of axions arising from the RR 4-form (and the universal axion), we can introduce a basis of 4-cycles $\Sigma_i$, $i=1,\ldots$, and define
\beqa
\phi_i\, =\, \int_{\Sigma_i} C_4
\eeqa
and let $\phi_0$ denote the universal axion. In this setup, the instanton generating (\ref{dbrane-inst-pot}) is an euclidean D3-brane wrapped on the 4-cycle $\sum_i n_i \Sigma_i$, and with a worldvolume gauge background with second Chern class $n_0$. 

The close resemblance of (\ref{dbrane-inst-pot}) and (\ref{multi-axion-grav-inst}), suggests the gravitational instantons may correspond to an effective description of non-perturbative effects which are microscopically described in terms of D-brane instantons. This would be much in the spirit of the correspondence between gravitational black hole solutions and systems of D-branes. In the following we provide some hints supporting this interpretation, at the semi-quantitative level (which is presumably the best one can hope for, in these non-supersymmetric setups).

An interesting observation is that BPS euclidean D-brane instantons (in general spacetime dimensions) can be described as solution to the euclidean supergravity equations of motion. These solutions source only the gravitational field and the axion(s) (complexified by a dilaton-like partner), have precisely the axionic charge structure of our solutions, and moreover in the string frame they are described by wormhole geometries, see e.g. \cite{Gibbons:1995vg,Bergshoeff:1998ry}, see also \cite{ Bergshoeff:2004fq,Bergshoeff:2004pg}. It is tantalizing to propose that our wormhole solutions correspond to a low-energy version of the D-brane instanton wormhole solutions, once the dilaton-like scalar partners are removed (by a hypothetical integrating out mechanism once these fields are made massive in the string setup). Notice that in the absence of dilation-like scalars, the string and Einstein frames are equivalent, so the wormhole nature of the D-brane instanton gravity solution becomes more physical.

An interesting hint in this direction comes from the dependence of the instanton actions on $g_s$. Let us consider the single axion case for simplicity. For gravitational instantons, the dependence of the action (\ref{instac1}) on $g_s$ is hidden in the 4d quantities $M_P$ and $f$. To make it explicit, let us consider the case of an axion from the RR 4-form, and perform the KK reduction of the 10d action for gravity and the RR 4-form, which (modulo the familiar self-duality issues) has the structure
\beqa
S_{10d} \, \sim\, \frac{1}{\alpha'{}^4}\int d^{10}x \sqrt{-G} \, (e^{-2\Phi} R\, +\, |F_5|^2)
\eeqa
with $\Phi$ the 10d dilaton. Defining the axion by $C_4=\phi \, \omega_4$ for some harmonic 4-form, in 4d we get
\beqa
S_{4d}\, \sim\,  \int d^4x \left(\, M_P^2\, \sqrt{-g}\, R \,+\, f^2\, |d\phi|^2 \,\right)
\eeqa
with 
\beqa
M_P^2\simeq \frac{V_6}{\alpha'^4  g_s^{\,2}} \quad ,\quad f^2\simeq \frac{1}{\alpha'^4}\int_{\IX_6} |\omega_4|^2
\eeqa
Hence the charge-$n$ gravitaional instanton action has a parametric dependence
\beqa
S_E\sim \frac{M_P\, n}{f}\sim \frac{n}{g_s}
\eeqa
where in the second equation we have ignored the $\omega_4$-dependent prefactor.

The above expression  has the precise parametric dependence expected for a charge-$n$ euclidean D-brane instanton (namely, a D3-brane multiply wrapped $n$ times on the corresponding 4-cycle). We take this agreement as a suggestive indication of the above indicated relation between gravitational instantons and D-brane instantons.

This analysis for the dependence on $g_s$ generalizes straightforwardly to the multi-axion case, using the action (\ref{instac2}). The agreement on the parametric dependence is in general not exact, but holds for large charges as follows. Consider  an euclidean D-brane with axion charge vector of the form $\vec{n}=p\vec{n_0}$, i.e. a large multiple of a basic vector $\vec{n_0}$. On general grounds, we expect the D-brane instanton action to scale as $S_{\rm D-inst}\sim p\, S_0$, with $S_0$ the action of the D-brane instanton associated to $\vec{n}_0$. The parametric dependence on $p$ is straightforwardly reproduced from (\ref{instac2}).

In some particularly simple instances, the action of D-brane instantons agrees with the precise dependence of the charges in  (\ref{instac2}), as we describe in a particular setup in the next section.

\subsection{Charge dependence in flat multi-axion setups}
\label{multi-flat}

In some very specific models, the specific dependence of the gravitational instanton action \eqref{instac2} on the charges of the axion lattice can be shown to survive the embedding into string theory.

Consider an euclidean $(p-1)$-brane wrapping a $p$-cycle $\mathcal{C}$ of the compactification manifold. Even if the brane is not BPS, one may estimate the action of the instanton via the usual Dirac-Born-Infeld action. In the particular case there are no gauge or $B$-fields turned on the brane worldvolume, this is just
\begin{align}S_{DBI}\approx\frac{\mu_p}{g_s}\int_\mathcal{C} dV_p=\frac{\mu_p}{g_s}\ V_p,\label{DBI}\end{align} 
proportional to the volume of the $p$-cycle. We will now see that, under some (admittedly very specific) circumstances, the action \eqref{DBI} has a dependence on the charges of the axion lattice similar to that of \eqref{instac2}.

In order to do this, it will be convenient to have an alternative expression for the volume element $dV_p$ of the brane. Let us assume that a globally defined $(n-p)$ form exists such that $* \omega_{n-p}= \lambda dV_p$, where $\lambda$ is a constant and $n$ is the dimension of the ambient space (typically $n=6$, although we may also consider submanifolds of the compactification manifold if we like). That is, $*\omega_{n-p}$ is proportional to the volume form of the $p$-cycle. Using the condition $dV_p\wedge *dV_p=dV_n$, one gets $\lambda=\sqrt{*(\omega_{n-p}\wedge* \omega_{n-p})}$, and thus we may write
\begin{align}
dV_p=\frac{*\omega_{n-p}}{\sqrt{*(\omega_{n-p}\wedge* \omega_{n-p})}}.
\end{align}
The usefulness of the above expression depends largely on the possibility of finding a suitable $\omega_{n-p}$. This is a hard thing to do in general, since $\omega_{n-p}$ cannot be guaranteed to exist, but it is a particularly simple thing to do for tori. In what follows, we will take the ambient space to be $T^n$, the $n$-dimensional torus, and the $p$-cycle will be a flat $p$-dimensional hiperplane.

As discussed above, an easy way to obtain axions in the effective theory coming from this string compactification is via the zero modes of RR $p$-forms. Namely, one expands
\begin{align}C_p=\sum \phi_i \beta_i\end{align}
with $\{\beta_i\}$ being the harmonic representatives  of a basis of $H_p(T^n, \IZ)$. Grouping the axions into a vector $\vec{\phi}$, their kinetic term will be of the form $\vec{\phi}^T\, \mathcal{G} \vec{\phi}$, with 
\begin{align} 
\mathcal{G}_{ij}=\langle \beta_i,\beta_j\rangle=\int_{T^n} \beta_i \wedge * \beta_j.\end{align}
The instantons coupled to these axions will be euclidean $(p-1)$-branes wrapped on cycles of $T^n$. $H^p(T^n,\IZ)$ and $H_{n-p}(T^n,\IZ)$ are naturally isomorphic via Poincar\'e duality. There is a single flat hyperplane in each class of $H^p(T^n,\IZ)$, and these may be expanded as $\mathcal{C}=\sum n_i\tilde\alpha_i$, where  $\{\tilde\alpha_i\}$ is the basis of $H^p(T^n,\IZ)$ dual to $\{\beta_i\}$. Then
\begin{align}\omega_{n-p}=\sum n_i \alpha_i\end{align}
satisfies that $*\omega_{n-p}\propto dV_p$. Here, $\alpha_i$ is the (unique) covariantly constant representative of the image of $\tilde\alpha_i$ via Poincar\'e duality. Furthermore, the $\alpha_i$ are harmonic. This means that the action of the instantons is, in this case
 \begin{align}S_{DBI}&\approx\int_\mathcal{C} \frac{*\omega_{n-p}}{\sqrt{*(\omega_{n-p}\wedge* \omega_{n-p})}}=\int_{T^n}\frac{*\omega_{n-p}\wedge \omega_p}{\sqrt{*(\omega_{n-p}\wedge* \omega_{n-p})}}\nonumber\\&=\sqrt{V_{T^n}}\sqrt{\int_{T^n}\omega_{n-p}\wedge* \omega_{n-p}},\label{DBIcomp}\end{align}
 where the volume factor arises because $*(\omega_{n-p}\wedge* \omega_{n-p})$ is constant, and hence 
 \begin{align}*(\omega_{n-p}\wedge* \omega_{n-p})=\frac{1}{V}\int_{T^n}\omega_{n-p}\wedge* \omega_{n-p}.\end{align}
 However, we now have
 \begin{align}\int_{T^n}\omega_{n-p}\wedge* \omega_{n-p}=\mathcal{G}^{-1}_{ij} n^i n^j,\end{align}
 since $\langle \alpha_i,\alpha_j\rangle$ is the inverse matrix to $\langle \beta_i,\beta_j\rangle$ by virtue of the duality between the two bases.

In this way, the dependence of the gravitational instanton action on its axionic charges is reproduced in the stringy setup. Strictly speaking however this result only holds for a torus. Exact computation of the action of euclidean non-BPS D-brane instantons in a generic Calabi-Yau manifold is a challenging matter into which we will not delve. 

One expects the gravitational instanton solution to be approximately valid when the charges are large and the curvature at the throat of the wormhole is low. If we fix a particular $p$-cycle and wrap $n$ branes around it, the action \eqref{DBIcomp} scales with $n$, as it did in the single-axion case, and at the end of the previous section. 

Notice that to recover the precise matching of the charge dependence it is crucial that the internal geometric objects are actually constant, so that the integrands are constant and integration amount to picking up volume factors. In this sense, the gravitational instanton can be regarded in the general case as an average assuming the internal geometric information is truncated to constant pieces.

\subsection{D-brane instantons and alignment}
\label{string-align}

Let us now apply the D-brane instanton viewpoint to the analysis of string theory realizations of transplanckian alignment models. In particular, we now show that string realization of the lattice alignment models described in section \ref{multiple} necessarily contain D-brane instantons which spoil the naive transplanckian inflaton field range allowed by the gauge non-perturbative axion potentials. In other words, the D-brane instantons take up the job carried out by the gravitational instantons in the  description in earlier sections.

Given the close relation between the axion charges of D-brane instanton and gravitational instanton, the analysis is simple. For a lattice alignment model, we recast it in the form shown in Figure \ref{figknp}, and consider D-brane instantons corresponding to the charges $(n_1,n_2)=p(1,-1)$, which in fact is of the form considered above for  D-brane instantons in multi-axion setups. This effectively corresponds to a single-axion model with a transplanckian decay constant $f$, so that the D-brane instanton action 
\beqa
S_{\rm D-inst}\sim \frac p{g_s}\sim \frac{p\, M_P}{f}
\eeqa
becomes small for any point inside the ellipse.

It is illustrative to consider a concrete setup. Several attempts to embed these alignment models in string theory are based on considering two stacks of D7-branes wrapped (possibly a multiple number of times $n$) on homologous 4-cycles (i.e. in the same class $[\Sigma_4]$), and carrying worldvolume gauge backgrounds with slightly different instanton numbers, $k$, $k'$. In terms of the homology class of the 4-cycle and the class of the point, the non-perturbative effects on these gauge sectors correspond to D-brane instantons with charge vectors $\vec{v}_1=(n,k)$, $\vec{v}_2=(n,k')$. Actually, we are interested in considering the lattice generated by these vectors, namely $n_1\vec{v}_1+n_2\vec{v}_2$, i.e. we use the potential to define the periodicities of the axions (using the more refined lattice defined by the homology amounts to including more instanton effects, but these do not reduce the field range, as explaine at the end of section \ref{sec:impli-multi}). The relevant D-brane instantons are given by $n_1=-n_2$, namely are of the form $p(\vec{v_1}-\vec{v}_2)$ which in terms of the underlying homology lattice corresponds to $(0,p(k-k'))$. In other words, the D-brane instanton is, at the level of charges, a superposition of $p$ D3-branes on $\Sigma_4$, each with  gauge instanton number $k$, and $p$ anti-D3-brane on $\Sigma_4$ with worldvolume instanton number $k'$. The lowest-energy state in this charge sector is a set of $p(k-k')$ D$(-1)$-brane instantons, whose action is essentially given by $p(k-k')M_P/f$.

The bottomline is that in string theory D-brane instantons provide the microscopic description of the effects which have been interpreted in terms of gravitational instantons in the effective theory in the earlier sections.

\section{Conclusions}
\label{conclu}

In this paper we have analyzed certain quantum gravitational effects on the possible transplanckian field ranges in single and multiple axion models (excluding setups with axion monodromy), and their implications for large field inflation model building. Concretely we have focused on gravitational instantons of the effective  theory of gravity coupled to the axion, which induce corrections to the axion potential. The effective theory with these degrees of freedom is actually the right framework, as they should be the only relevant ones in inflationary applications. In most cases, our results are in agreement with the Weak Gravity Conjecture, although the existence of the euclidean instantons considered in this paper is in principle independent of the validity of the WGC.

Our conclusions are as follows. In  single-axion models, gravitational instanton effects spoil transplanckian axion decay constant $f$, since the instanton action decreases with $f$. In multiple axion setups, we have considered different scenarios of axion alignment. We have shown that in models of lattice alignment, the parametrically transplanckian axion suffers gravitational instanton effects which jeopardize its transplanckian field range, so in this sense there is not advantage with respect to the single field case. Finally, we have shown that kinetic alignment can be used to achieve moderately transplanckian field ranges, of the form $\sqrt{N}f_{\rm max}$, where $N$ is the number of axions and $f_{\rm max}$ the maximum axion periodicity, corresponding to travelling along a diagonal in the lattice. In the effective single-axion model along this transplanckian direction, the gravitational instantons which would spoil the transplanckian range are absent due to a $\IZ_N$ discrete gauge symmetry.

The last result suggests a generalization of the Weak Gravity Conjecture, in theories with appropriate discrete gauge symmetries. This generalization applies not only to axions, but to general theories with gravitational and gauge interactions for charged extended objects of different dimensions (by invoking generalized discrete gauge symmetries, for which charged objects are extended as well).

Our techniques and results are useful in improving large field inflation model building with axions, a most relevant topic given the near-future prospects of exploration of B-mode polarization in the CMB \cite{Creminelli:2015oda}. In addition, they are an interesting further step in the study of the validity of effective field theory in quantum gravity theories. We hope our work triggers further research in these two fronts. 

\section*{Acknowledgments}

We thank Luis Ibanez, Fernando Marchesano, Pablo Soler, Gianluca Zoccarato and Mario Herrero-Valea for useful discussions. We thank Ander Retolaza for very  illuminating discussions during the early stages of this work. This work is partially supported by the grants  FPA2012-32828 from the MINECO, the ERC Advanced Grant SPLE under contract ERC-2012-ADG-20120216-320421 and the grant SEV-2012-0249 of the ``Centro de Excelencia Severo Ochoa" Programme. M.M. is supported by a ``La Caixa'' Ph.D scholarship. I.V. is supported through the FPU grant AP-2012-2690.

\bibliographystyle{unsrt}
\bibliography{transplanckian}

\end{document}